\documentclass[showpacs,showkeys,reprint,onecolumn,longbibliography]{revtex4-1}
\usepackage{amssymb}
\usepackage{amsmath}
\usepackage{graphicx}
\usepackage{amsbsy}
\usepackage{float}
\usepackage{color}

\newcommand{\bq}{\begin{eqnarray}}
\newcommand{\eq}{\end{eqnarray}}
\newcommand{\bqn}{\begin{eqnarray*}}
\newcommand{\eqn}{\end{eqnarray*}}
\newcommand{\xx}{{\bf x}}
\newcommand{\yy}{{\bf y}}
\newcommand{\zz}{{\bf z}}
\newcommand{\rr}{{\bf r}}
\newcommand{\RR}{{\bf R}}

\newcommand{\qq}{{\bf q}}
\newcommand{\pp}{{\bf p}}
\newcommand{\vv}{{\bf v}}
\newcommand{\QQ}{{\bf Q}}
\newcommand{\PP}{{\bf P}}
\newcommand{\VV}{{\bf V}}

\newcommand{\an}{}
\newcommand{\ann}{}
\newcommand{\annn}{}
\newcommand{\joris}{}

\begin{document}
%%%%%%%%%%%%%%%%%%%%%%%%%%%%%%%%%%%%%%%%%%%%%%%%%%%%%%%%%%%%%%%%%%%%%%%%%%%%%%%
%%%%%%%%%%%%%%%%%%%%%%%%%%%%%%%%%%%%%%%%%%%%%%%%%%%%%%%%%%%%%%%%%%%%%%%%%%%%%%%
%%%%%%%%%%%%%%%%%%%%%%%%%%%%%%%%%%%%%%%%%%%%%%%%%%%%%%%%%%%%%%%%%%%%%%%%%%%%%%%
\title{The structure of colloidosomes with tunable particle density:
  simulation {\sl vs} experiment}  

\author{Riccardo Fantoni}
\email{rfantoni27@sun.ac.za}
\affiliation{National Institute for Theoretical Physics (NITheP) and
Institute of Theoretical Physics,  University of
Stellenbosch, Stellenbosch 7600, South Africa} 
\author{Johannes W. O. Salari}
\email{jorissalari@gmail.com}
\affiliation{Department of Polymer Chemistry, University of Technology
Eindhoven, P.O. Box 513, 5600 MB Eindhoven, the Netherlands}
%\author{Hans M. Wyss}
%\email{H.M.Wyss@tue.nl}
%\affiliation{Institute for Complex Molecular Systems and Department of
%Mechanical Engeneering, P.O. Box 513, 5600 MB Eindhoven, the Netherlands}
\author{Bert Klumperman}
\email{bklump@sun.ac.za}
\affiliation{Department of Polymer Chemistry, University of Technology
Eindhoven, P.O. Box 513, 5600 MB Eindhoven, the Netherlands and
Stellenbosch University, Department of Chemistry and Polymer
Science, Private Bag X1, 7602 Matieland, South Africa}

\date{\today}

\begin{abstract}
Colloidosomes are created in the laboratory from a Pickering
emulsion of water droplets in oil. The colloidosomes have 
approximately the same diameter and by choosing (hairy) particles of
different diameters it is possible to control the particle density on
the droplets. The experiment is performed at room temperature. The
radial distribution function of the assembly of (primary) particles on
the water droplet is measured in the laboratory and in a computer
experiment of a fluid model of particles with pairwise interactions on
the surface of a sphere.
\end{abstract}


\maketitle
%%%%%%%%%%%%%%%%%%%%%%%%%%%%%%%%%%%%%%%%%%%%%%%%%%%%%%%%%%%%%%%%%%%%%%%%%%%%%%%
\section{Introduction}
%%%%%%%%%%%%%%%%%%%%%%%%%%%%%%%%%%%%%%%%%%%%%%%%%%%%%%%%%%%%%%%%%%%%%%%%%%%%%%%
Colloidosomes are hollow spherical structures that are formed by the
assembly of colloidal particles {\annn{at} the interfaces of two
immiscible liquids} \cite{Dinsmore2002,*Zeng2006,*Subramanian2005}.
\annn{As a result the particles are arranged in a shell that is inherently
porous.}  

The assembly of \annn{colloidal} particles at liquid interfaces is used in
various applications
\cite{Lee1999,*Dickinson2010,*Rousseau2009,*Lu2005,*Frelichowska2009,*Hansen2001}. 
Moreover, it is a promising technique for the synthesis of novel
materials \cite{Binks2006,*Hong2006} and has recently 
led to the development of colloidosomes \cite{Dinsmore2002},
nano-composite particles \cite{Bon2007}, porous solids
\cite{Pitard1995}, and foams \cite{Choen-Addad}.   

In this work, we study \joris{colloidosomes that are composed} of
uncharged spherical polystyrene particles of $\mu$m size moving on the
surface of a water droplet in oil. Similar studies have also been done
with charged particles 
\cite{Pieranski1980,*Aveyard2002,*Leunissen2007,*Masschaele2010,*Guzowski2011}.
The study of particles on the surface of a 
sphere dates back to the old Thomson problem
\cite{Bowick2002,*Kiessling2009} for classical electrons. The
statistical physics problem of a one component plasma on a sphere
has been solved exactly analytically at a special value of the
temperature \cite{Caillol1981}. Non point-wise particles on a sphere
have the additional complication of the geometrical frustration, which
can be described through the so called grain boundary scars
\cite{Bausch2003,Lipowsky2005,Bowick2000}. There have been attempts to
formulate a statistical geometry of particle packing
\cite{Sastry1997}. These systems opened up a new field of research
that studies the effect of curvature and topology of various surfaces
on the organization of matter in a more general sense
\cite{Bowick2009}. Structuring at the surface of a droplet 
can be viewed as a two dimensional analog to fluid like behavior,
crystallization, or glass formation in three dimensional systems
\cite{Likos2001,*Zaccarelli2007}.   

Fluids on Riemannian surfaces have been the subject of various
studies with few exact analytical results
\cite{Caillol1981,Fantoni2003,Fantoni2008}, some
approximate theories
\cite{Sausset2010,*Paez2003,*Mendez2008},
and \ann{many} Monte Carlo (MC) simulations
\cite{Prestipino1992,*Prestipino1993}.

Colloidosomes with tunable particle density were synthesized
experimentally. \cite{Salari2010,*Salari-Thesis}. A sintering procedure is
then used to create capsules which can be easily handled. The capsules
are then dried to obtain {\sl colloidal cages}. 
\joris{The synthetic details are explained in the next section.}

In this work we give the simplest statistical physics
description of the colloidosome, where we describe the interaction of
the colloidal particles with the surrounding media, water and oil,
simply as a holonomic constraint on the particles positions to stay
\ann{at} the 
water-oil (w/o) interface and treat them as a {\sl fluid} of a fixed
number of particles  
moving on a sphere, the droplet of water in oil, with a mutual 
pairwise interaction, the {\sl pair-potential}, at a temperature
$T$. Additional frictional effects have been \annn{neglected}
\cite{Smith2010,*Bruneau2002}. \annn{The} assembly of 
particles on \annn{the} sphere is studied \annn{both in the laboratory
and with a computer experiment under certain conditions: number
density and temperature.} The structural   
arrangement of the particles is characterized through the radial
distribution function. \annn{The colloidal particles created in our
  laboratory are polystyrene solid spherical hairy particles with
  controllable diameter of the order of $3\,\mu$m. The particles will
  then exhibit a hard core interaction.} \joris{Two types of particle
  pair-potentials were used in the Monte Carlo simulation
  of the fluid, namely the hard-sphere one
  and the polarizable hard-sphere one.}  

The work is organized as follows: in Section \ref{sec:model} the
colloidosome is described; in Section
\ref{sec:gr} the radial distribution function as a means to probe the
structure of the colloidosome is presented in its mathematical
definition described in Appendix \ref{app:1}, its MC estimator, and
its experimental measure; in 
Section \ref{sec:results} the MC simulation results are presented; in
Section \ref{sec:exp} the theoretically exact
results of the MC simulation and the experimental results are
compared; Section \ref{sec:conclusions} is devoted to concluding
remarks.  

%%%%%%%%%%%%%%%%%%%%%%%%%%%%%%%%%%%%%%%%%%%%%%%%%%%%%%%%%%%%%%%%%%%%%%%%%%%%%% 
\section{The experimental system {\sl versus} the statistical physics problem}
%%%%%%%%%%%%%%%%%%%%%%%%%%%%%%%%%%%%%%%%%%%%%%%%%%%%%%%%%%%%%%%%%%%%%%%%%%%%%%
\label{sec:model}

The details for the synthesis of the colloidosomes can be found in
our previous work \cite{Salari2010,*Salari-Thesis}. Working
at room temperature, we first disperse the colloidal particles in a
hydrocarbon oil (heptane). Then, water is added while the solution is being
stirred vigorously. The function of the shear is two fold. It causes
the water to break up into small water droplets and at the same time
it allows to overcome the barrier for adsorption of the particles
which are assembled randomly at the w/o interface of the
droplets. Eventually a stable Pickering emulsion \cite{Pickering1907}
of water droplets covered by polystyrene (pS) particles, the
colloidosomes, in oil is formed. \annn{The colloidosomes formed have
  all approximately the same 
  diameter, and this, as well as the number of colloidal particles on
  each colloidosome, does not change after the colloidosomes are
  formed.} For further imaging with scanning electron microscopy (SEM) \joris{the particles surrounding the droplets need to be (partially) sintered in order to form a continuous and stable
shell around the droplet, \annn{a capsule}.}
Heating the
mixture to 35 $^\circ$C for 30 minutes proved sufficient to do so. 
\joris{A small amount of the sintered colloidosome dispersion is
  placed on the SEM sample 
  holder} and dried in the fumehood. This removes both the water and
  heptane remnants and leaves only the capsules, the
  {\sl colloidal cages}, which can then be imaged by SEM. 

The colloidal polystyrene solid particles are synthesized by the
dispersion polymerization of styrene in alcohol/water
\cite{Salari2010,*Salari-Thesis}, which is a 
well established technique for the formation of highly uniform polymer
particles with a narrow size polydispersity. AIBN
(azo-bis(isobutyronitrile)) is used as the initiator. The presence of a
polymeric stabilizing surface functional group is required for a
controlled synthesis of the particles. \joris{A non ionic polymeric
stabilizer (poly(N-vinylpyrrolidone) (pVP)) is used for this
study. Hence, there is no charge on the surface of the particles.}
During the polymerization, pVP, which efficiently 
adsorbs on polystyrene, is attached to the particle. In ethanol and
water, pVP is soluble. The polymeric chains are extended (with a radius
of gyration of $R_g \approx 15\, \text{nm}$) and are responsible
for steric repulsion as two particles get in close contact. This
guarantees the steric stabilization of the colloidal suspension. In
the oil phase, pVP is  
insoluble and the polymer chains are collapsed on the surface of the
particle, \joris{resulting in an attractive potential among the particles.} 
The final dispersion in ethanol/water, therefore, consists of
polystyrene particles 
that are sterically stabilized with a layer of pVP. During the
polymerization of styrene, pVP attaches to the particle by both
physical adsorption to the particle's surface and chemical
grafting. These two mechanisms occur simultaneously, however it is not
known to which extent. It is believed that the predominant mechanism
for stabilization is the physical adsorption. The particles are
washed with pure ethanol by three centrifugation/redispersion cycles
in order to remove the residual physically adsorbed pVP. The particles
settle due to the  
centrifugation. The supernatant solution is decanted and clean ethanol
is added to the remaining particles. The particles are then
redispersed and the whole procedure is repeated 3 times. The
physically adsorbed pVP is removed, which produces a lower
colloidal stability of the particles in ethanol. Large aggregates were
observed during this procedure, which is an indication for the
presence of an attractive component in the pair-potential between the
colloidal particles in the suspension. The chemically grafted pVP
remains attached to the particle's surface. The remaining particles
are dried and redispersed in heptane, before the colloidosomes are
synthesized. 
In conclusion, the surface chemical properties of the particles are
mainly determined by polystyrene and the fraction of pVP that is
chemically grafted to the surface, although a precise estimate of the
grafting density is
lacking. It is believed that the grafting density is low, due to the
poor colloidal stability of the particles in ethanol after removing
the physically adsorbed pVP.

\begin{figure}[H]
\begin{center}
\includegraphics[width=6.5cm]{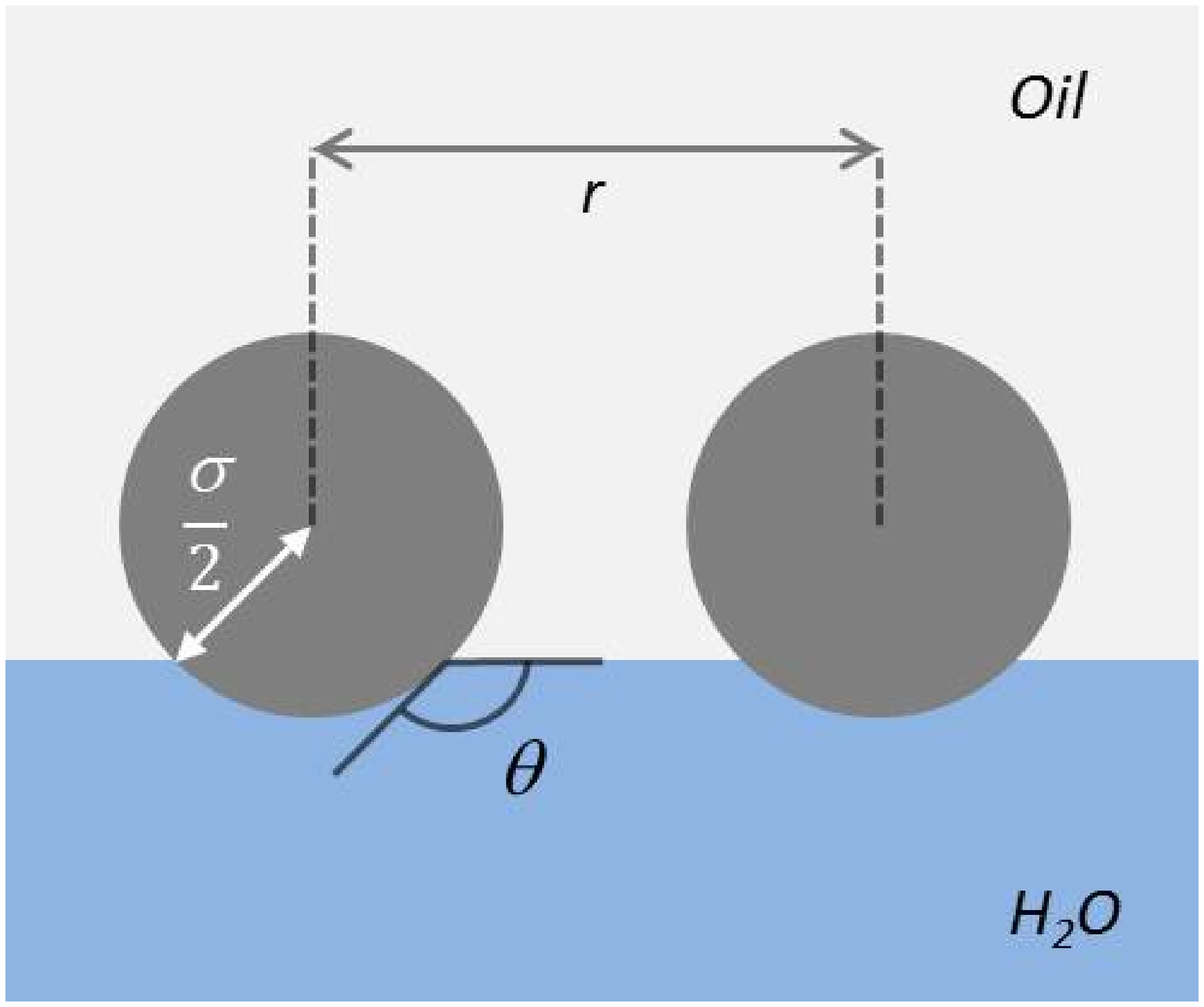}
\hspace{.5cm}
\includegraphics[width=7cm]{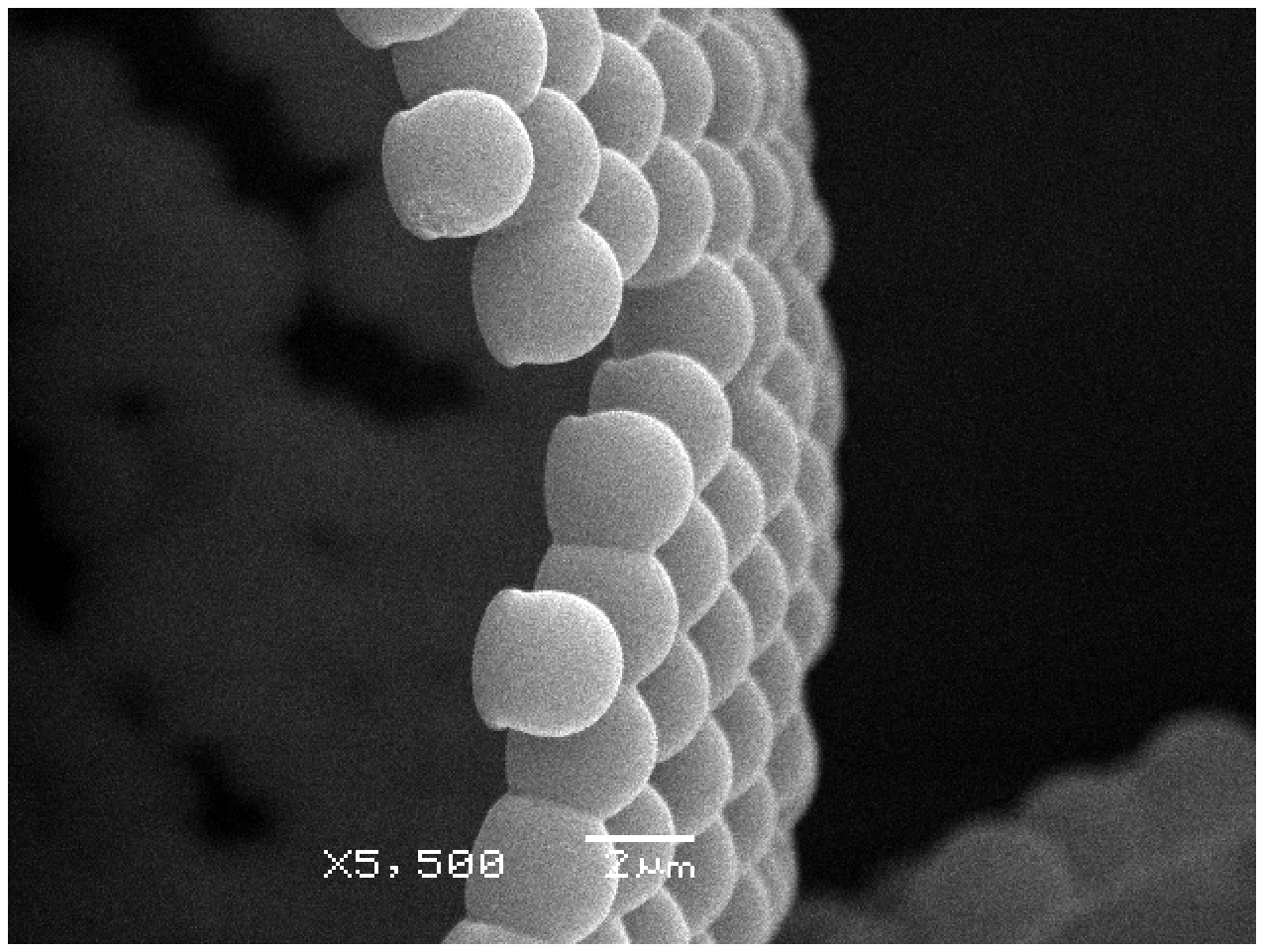}
\end{center}
\caption{\annn{The left panel shows a} schematic representation of the
  equilibrium position of the 
particles at the w/o interface; $\theta$ is the
three phase contact angle, $r$ is the particle separation, and
$\sigma$ is the particle diameter. \annn{The right panel shows} a side
view of the particle shell 
of a colloidosome made with scanning electron microscopy.} 
\label{fig:2}
\end{figure}

In Pickering emulsions, the particle is adsorbed at the w/o
interface and is partly immersed in the oil and water phase. The
extent of immersion in both phases will eventually have an influence
on the particle pair-potential. The three phase contact angle $\theta$
is used to denote the position at the interface as shown in 
Fig. \ref{fig:2} \footnote{
The right panel of the Fig. \ref{fig:2} shows clearly flattening
of the particles on the inside of the capsules. We are convinced
that this is an artifact of the sintering process. During this process
flattening of the particles occurs, which we ascribe to particle
deformation to reduce the contact area between particle and
water.}. It was determined in our earlier work 
\cite{Salari2010,*Salari-Thesis} and is approximately
$\theta\approx 130\ensuremath{^\circ}$, which means that the particle is
predominantly immersed in the oil. The surface tension of
the particle is altered by the presence of the surface stabilizing
groups, which affects the wetting properties and the equilibrium
position of the particles at the interface \cite{Salari2011}.  
An atomistic \cite{Landau3} level
of description of the core of the solid hairy particles suggests the
use of Hamaker \cite{Hamaker1937} calculation for the determination of
the interaction between the two spherical cores. The calculation
predicts an attractive pair-potential which, neglecting the detailed
behavior close \annn{to} contact, is proportional to $(\sigma/r)^6$, $r$ being
the distance between the centers of the two cores of diameter
$\sigma$. Unlike this attraction, which is always present, the steric
repulsion will have a very small range when $\theta$ is obtuse since
in this case the particles contact occurs in the oil phase
\cite{shs}. We thus 
expect the balance between the attractive interaction and the steric
effects to depend on the angle $\theta$. Moreover other kinds of
interactions such as depletion, hydrophobic, solvation, or capillary
should be taken into account for an accurate description of our
system. The simplest description for the pair-potential between the
particle is the hard-spheres one.

The surface of a sphere of diameter $D=2R$ is $A=\pi D^2$. The surface
area that a particle, with diameter $\sigma$, can occupy is 
approximately $a=\sqrt{3}\sigma^2/2$. The maximum
number of particles that can pack the surface of a sphere is
\annn{approximately} 
\bq
N_{max}\lesssim\frac{A}{a}
=\frac{2\pi}{\sqrt{3}}\left(\frac{D}{\sigma}\right)^2~,
\eq
where we assumed that the particles are in a close packing
regular hexagonal lattice neglecting curvature effects. The
maximum reduced particle density on the sphere will then be
$\rho_{max}\sigma^2\lesssim 2/\sqrt{3}\approx 1.155$.

\ann{Similar experiments \cite{Jiang2008} make use of water and a
  liquid of higher density, for the initial solution of the two
  immiscible liquids. The droplets in the emulsion will now be of the
  higher density liquid. In the limit of \annn{droplets of}
  very high density the particles are expected to be essentially
  \annn{unable to move on the droplet. Our working hypothesis will be,
    instead, to consider the particles as moving freely on the droplet
    surface, completely neglecting the presence of the solvent}.} 
We then treat the colloidosome of diameter $D$, number of particles
$N$, and temperature $T$, through a canonical ensemble classical
statistical physics description \annn{of} the assembly of particles
on the water droplet as a fluid of particles constrained to move on
the surface of a sphere with a pairwise interaction, the
pair-potential.

%%%%%%%%%%%%%%%%%%%%%%%%%%%%%%%%%%%%%%%%%%%%%%%%%%%%%%%%%%%%%%%%%%%%%%%%%%%%%%%
\subsection{The pair-potential}
%%%%%%%%%%%%%%%%%%%%%%%%%%%%%%%%%%%%%%%%%%%%%%%%%%%%%%%%%%%%%%%%%%%%%%%%%%%%%%%
\label{sec:pp}

Fixing the pair-potential completely defines the {\sl fluid model},  
as described in Appendix \ref{app:1} and Eq. (\ref{tpot}).
 
The simplest interaction between two colloidal particles is
the hard-spheres (HS) pair-potential
\bq
\phi_\text{HS}(r)=\left\{
\begin{array}{ll} \displaystyle
+\infty & r<\sigma\\ \displaystyle
0       & r>\sigma
\end{array}
\right.~,
\eq
where $\sigma$ is the diameter of the spheres and $r$ is the Euclidean
center to center distance (see Eq. (\ref{ed})).

The interaction between two neutral particles far apart is dominated
by dipolar forces. The simplest model potential, suggested by the
London forces \cite{Gazzillo2006},
corresponds to hard-spheres of diameter $\sigma$ 
with dispersion attractions, \annn{the} polarizable hard-spheres (PHS),
\bq
\phi_\text{PHS}(r)=\left\{
\begin{array}{ll} \displaystyle
+\infty & r<\sigma\\ \displaystyle
-\epsilon_\text{PHS}\left(\frac{\sigma}{r}\right)^6 & r>\sigma
\end{array}
\right.~,
\eq
where $\epsilon_\text{PHS}=A^H/36$ is a positive energy proportional to
Hamaker constant \cite{Hamaker1937} $A^H$ which is a property of the
material of which the particles are made and of the environment where
the particles are immersed. We here neglect the close to contact $-A^H
\sigma/24/(r-\sigma)$ Hamaker behavior as on the average, even in a
closed packed configuration, there are many more pair interactions
between far particles than between close ones.

%%%%%%%%%%%%%%%%%%%%%%%%%%%%%%%%%%%%%%%%%%%%%%%%%%%%%%%%%%%%%%%%%%%%%%%%%%%%%% 
\section{The radial distribution function}
%%%%%%%%%%%%%%%%%%%%%%%%%%%%%%%%%%%%%%%%%%%%%%%%%%%%%%%%%%%%%%%%%%%%%%%%%%%%%% 
\label{sec:gr}

In this work we probed the structure of the colloidosome using the
radial distribution function (RDF). We compare the experimental RDFs
with the ones obtained from MC simulations of a fluid of particles
moving on a sphere and interacting with a model
pair-potential of the kinds described in Section
\ref{sec:pp}. This procedure will allow us to determine which
interaction model best describes the experimental assembly of
particles. Choosing $\sigma$ as the unit length, the statistical
physics problem only depends on the number of particles $N$ and the
reduced density $\rho\sigma^2=N/[\pi(D/\sigma)^2]$ for the athermal HS
model and also on the reduced temperature $k_BT/\epsilon_\text{PHS}$
for the PHS one. 
 
%%%%%%%%%%%%%%%%%%%%%%%%%%%%%%%%%%%%%%%%%%%%%%%%%%%%%%%%%%%%%%%%%%%%%%%%%%%%%% 
\subsection{In the Monte Carlo simulation}
%%%%%%%%%%%%%%%%%%%%%%%%%%%%%%%%%%%%%%%%%%%%%%%%%%%%%%%%%%%%%%%%%%%%%%%%%%%%%% 
On a sphere, the Monte Carlo simulation \cite{Allen-Tildesley} solves
exactly the statistical physics problem as, since one does not have
the additional thermodynamic limit problem, it reduces to an
integration, as described in Appendix \ref{app:2}.

The particles positions are $\RR=(\rr_1,\rr_2,\ldots,\rr_N)$ with
\bq
\rr_i=R[\sin\theta_i\cos\varphi_i\hat{\xx}+
\sin\theta_i\sin\varphi_i\hat{\yy}+
\cos\theta_i\hat{\zz}]~.
\eq
The Euclidean distance between particles $i$ and $j$ is given by
\bq \label{ed}
r_{ij}=R\sqrt{2-2\hat{\rr}_i\cdot\hat{\rr}_j}~,
\eq
where $\hat{\rr}_i=\rr_i/R$ \annn{is the versor that from the center
of the sphere points towards the center of the $i$th particle}.

The density of particles on the surface of the sphere is 
\bq
\rho=\frac{N}{4\pi R^2}~.
\eq

In the MC simulation \cite{Allen-Tildesley} the \annn{RDF} between two
points on the sphere, $\rr$ and $\rr^\prime$, is calculated through
the following ``histogram'' estimator 
(see Eq. (\ref{gr}))
\bq
g(d)=\langle g^\text{histogram}(d,\RR)\rangle~,
\eq
where $d=2R\sin(\arccos(\hat{\rr}\cdot\hat{\rr}^\prime)/2)$ is the
Euclidean distance between $\rr$ and $\rr^\prime$, 
$\langle\ldots\rangle$ $=$ $\int_{S_R^N}\exp(-\beta U_N(\RR))$ $\ldots
d\RR$ $/$ $\int_{S_R^N}\exp(-\beta U_N(\RR))\,d\RR$ is the thermal average,
here 
\bq \label{tpot}
U_N(\RR)=\sum_{i<j}\phi(r_{ij})~,
\eq 
is the total potential energy of the fluid of particles, $\phi$ is the
pair-potential, and the integrals are taken in such way that
$\rr_i\in S_R$ for $i=1,2,\ldots,N$ with $S_R$ the sphere of diameter
$D=2R$, so that $d\RR=\prod_id\rr_i$ with $d\rr_i=R^2d\Omega_i=
R^2\sin\theta_id\theta_i d\varphi_i$, and
\bq
g^\text{histogram}(d,\RR)=\sum_{i\neq j}
\frac{1_{[d-\Delta/2,d+\Delta/2[}(r_{ij})}{Nn_{id}(d)}~
\eq
here $1_{[a,b[}(r)=1$ if $r\in[a,b[$ and 0
otherwise, and
\bq
n_{id}(d)=N\left[\left(\frac{d+\Delta/2}{2R}\right)^2-
\left(\frac{d-\Delta/2}{2R}\right)^2\right]~,
\eq
is the average number of particles on the surface
$[d-\Delta/2,d+\Delta/2[$ for the ideal gas of density $\rho$. $\rho^2g(d)$ 
gives the probability that sitting on a particle at $\rr$ one has to
find another particle at $\rr^\prime$. 

%%%%%%%%%%%%%%%%%%%%%%%%%%%%%%%%%%%%%%%%%%%%%%%%%%%%%%%%%%%%%%%%%%%%%%%%%%%%%% 
\subsection{In the experiment}
%%%%%%%%%%%%%%%%%%%%%%%%%%%%%%%%%%%%%%%%%%%%%%%%%%%%%%%%%%%%%%%%%%%%%%%%%%%%%%
 
The positional data of the particles in {the colloidal cages} is
directly extracted from SEM images, which allowed the calculation
of the particle separation for all visible particle pairs. $\Delta$ is
set to an arbitrary value of $\sigma/20$. To exclude edge effects, a
selection of particles located sufficiently at the center of
the SEM image of {the colloidal cage} \annn{is} taken into account. The
RDF is determined from just one hemisphere. The particle positions
from 5 \annn{SEM} 
images of similar {colloidal cages} were used for the statistical
average. The detailed procedure, the selection of particles, and
validation of the procedure is described in our previous work
\cite{Salari2010,*Salari-Thesis}. In
that work we calculated the radial distribution function from just one
SEM image. Here we refined that analysis averaging the results from 5
SEM images which is in spirit closer to the procedure used in the MC
simulations. Although five images are still a rather small number our
present procedure carry nevertheless more information than the one
used in paper \cite{Salari2010}. The absolute error on the
experimental $g(r)$ is around $0.3$. In the experimnt unlike in the
simulations each image measurememnt is scorrelated from the other.   

%%%%%%%%%%%%%%%%%%%%%%%%%%%%%%%%%%%%%%%%%%%%%%%%%%%%%%%%%%%%%%%%%%%%%%%%%%%%%% 
\section{Monte Carlo results}
%%%%%%%%%%%%%%%%%%%%%%%%%%%%%%%%%%%%%%%%%%%%%%%%%%%%%%%%%%%%%%%%%%%%%%%%%%%%%% 
\label{sec:results}
We performed constant $N$, $\rho$, and $T$ canonical MC simulations
\cite{Allen-Tildesley,*Frenkel-Smit}. A typical run would consist of
about $5\times 10^5N$ single particle moves, keeping the
acceptance ratios constant ($\approx 0.3$). In all the
presented graphs of the simulated RDF the statistical error from the
MC integration are not visible on the \annn{chosen} scale.

We initially chose the PHS model pair-potential to see how the RDF
would change upon changing the temperature and the density. \an{In
order to find agreement with the experimental results it proved
necessary to use the simpler HS model, as shown in Section
\ref{sec:exp}. We then compared the HS results with the soft-sphere
model
$\phi_\text{SS}(r)=\epsilon_\text{SS}\left(\frac{\sigma}{r}\right)^6$
considered in Ref. \onlinecite{Mendez2008}. For case ``a'' of Table
\ref{tab:exp} a reduced temperature of $k_BT/\epsilon_\text{SS}=0.05$
is sufficient to have similar RDFs for the HS and the SS model on
the half hemisphere, but when looking at the RDFs on the whole sphere,
the SS RDF, unlike the HS one, shows relevant correlations between
particles at opposite poles $(|g(2R)-1|\approx 0.3)$.}

%%%%%%%%%%%%%%%%%%%%%%%%%%%%%%%%%%%%%%%%%%%%%%%%%%%%%%%%%%%%%%%%%%%%%%%%%%%%%%%
\subsection{Dependence on the temperature}
%%%%%%%%%%%%%%%%%%%%%%%%%%%%%%%%%%%%%%%%%%%%%%%%%%%%%%%%%%%%%%%%%%%%%%%%%%%%%%%
The HS model is athermal so the structure is independent of
temperature but only depends on the density. We thus simulated the
colloidosome ``c'' in Table \ref{tab:exp} with the PHS model. And we 
chose different values of the reduced temperature,
$k_BT/\epsilon_\text{PHS}$, to see how the RDF would change. 

\begin{figure}[H]
\begin{center}
\includegraphics[width=12cm]{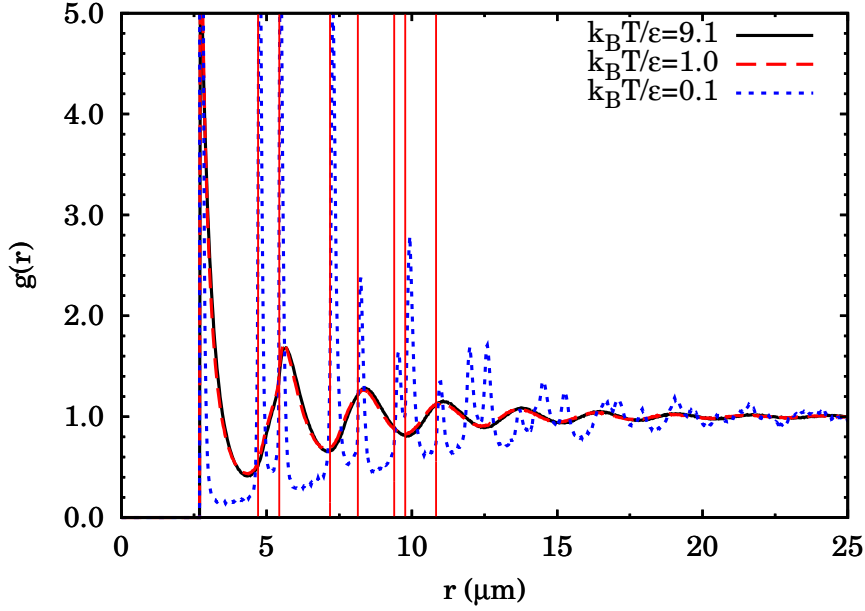}
\end{center}
\caption{(Color online) RDF for \annn{approximately} one hemisphere of
  the colloidosome 
  ``c'' of Table 
  \ref{tab:exp} with $N=1498$, $D/\sigma=23.8$ and various reduced 
  temperatures, calculated with MC simulations of the PHS fluid. Also
  shown are the locations (vertical lines) of 
  the first eight coordination shells of a regular planar hexagonal
  lattice of the hard core particles (here $\sigma=2.72\, \mu$m). The
  mismatch between the peaks of the RDF and these shells is a
  manifestation of the curvature of the surface.}  
\label{fig:grt}
\end{figure}

As expected we found the occurrence of an ordered structure at small
reduced temperatures (see Fig. \ref{fig:grt}). In particular we
observe the formation of a regular hexagonal lattice distorted by the
curvature of the spherical surface. Recall that in
a planar perfect hexagonal lattice arrangement of \annn{spheres} of diameter
$\sigma$ the first coordination shells are as follows:
$r/\sigma=1,\sqrt{3},2,\sqrt{7},3,2\sqrt{3},\sqrt{13},4,\sqrt{19},
\sqrt{21},5,\ldots$.   

From Fig. \ref{fig:grt} we can clearly see how at this reduced density,
$0.84$, well below the maximum density, the PHS model reduces to the
HS model for reduced temperatures $\gtrsim 1$. As we lower the
temperature, the attractive tail in the pair-potential starts to play a
role resulting in a solidification of the fluid. As the fluid
crystallizes, it may experience the cage effect going
through glassy phases. The particles become confined in
transient cages formed by their neighbors. This prevents them from
diffusing freely \annn{on the surface of the sphere} \cite{Weeks2002}. A
related problem is the extremely long MC equilibration
time necessary to draw the RDF of the figure at a temperature of 0.1,
starting from a disordered initial configuration. In the limit of
$T\to 0$, in our calculation, the 
equilibrium configuration of the (classical) particles is the {\sl one}
$\RR_0$ for which $U_N(\RR)$ has its minimum: the probability density
is zero everywhere except on $\RR_0$. We have a spontaneous breaking of
the rotational symmetry (see Appendix \ref{app:1}). The monotonously
increasing tails in the PHS pair-potential produce an equilibrium
configuration with the particles forming one cluster of touching
spheres. \an{On the contrary in the SS model the equilibrium
configuration will be one where the inter-particle spacing depends on
the density.}

%%%%%%%%%%%%%%%%%%%%%%%%%%%%%%%%%%%%%%%%%%%%%%%%%%%%%%%%%%%%%%%%%%%%%%%%%%%%%%%
\subsection{Dependence on the density}
%%%%%%%%%%%%%%%%%%%%%%%%%%%%%%%%%%%%%%%%%%%%%%%%%%%%%%%%%%%%%%%%%%%%%%%%%%%%%%%
For case ``c'' in Table \ref{tab:exp} ($D/\sigma=23.8$ and
$\sigma=2.72\,\mu$m) we chose different values of the 
density to see how the RDF would change for the PHS model at a
relatively high value of the reduced temperature
$k_BT/\epsilon_\text{PHS}=9.1$.  

\begin{figure}[H]
\begin{center}
\includegraphics[width=12cm]{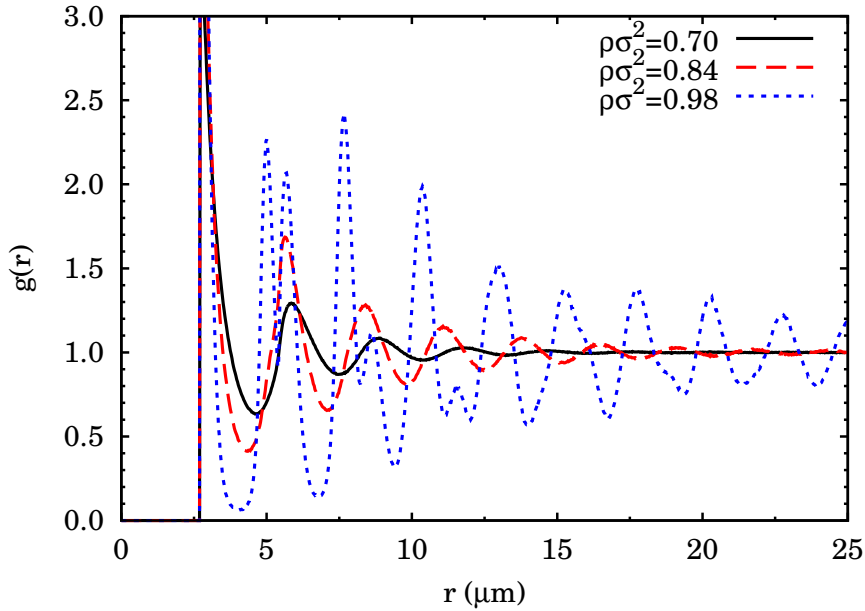}
\end{center}
\caption{(Color online) RDF for \annn{approximately} one hemisphere of
  the colloidosome with 
  $D/\sigma=23.8$ and $\sigma=2.72\,\mu$m at a reduced temperature 
  $k_BT/\epsilon_\text{PHS}=9.1$ and various densities, calculated
  with a MC simulation of the PHS fluid.}  
\label{fig:grd}
\end{figure}

We succeeded in reaching high particle densities (without overlaps)
by placing one particle at the north pole and then others centered at
$\theta=2n\arcsin(1/2R)$ and $\varphi=2m\arcsin(1/2R\sin\theta)$ with
$n,m=1,2,3,\ldots$. This way we were, in particular, able to reach the
$0.91$ critical density observed by Prestipino Giarritta {\sl et al.}
\cite{Prestipino1992,*Prestipino1993} for HS. In doing so we observed
the splitting of 
the second peak into a pair of adjacent peaks corresponding to the
second and third coordination shells of a regular hexagonal lattice
(see Fig. \ref{fig:grd}).

From Fig. \ref{fig:grd} we can clearly see how the fluid tends to
reach an ordered phase at high densities (even at high
temperatures). The realization of these ordered phases will go through
the formation of colloidal geometrical cages (due to geometrical
frustration) on the surface of the water droplet which is inevitable
as the density slowly approaches the maximum density at any temperature. 

%%%%%%%%%%%%%%%%%%%%%%%%%%%%%%%%%%%%%%%%%%%%%%%%%%%%%%%%%%%%%%%%%%%%%%%%%%%%%% 
\section{Comparison between experimental and Monte Carlo simulation results}
%%%%%%%%%%%%%%%%%%%%%%%%%%%%%%%%%%%%%%%%%%%%%%%%%%%%%%%%%%%%%%%%%%%%%%%%%%%%%% 
\label{sec:exp}

\begin{figure}[H]
\begin{center}
\includegraphics[width=7.5cm]{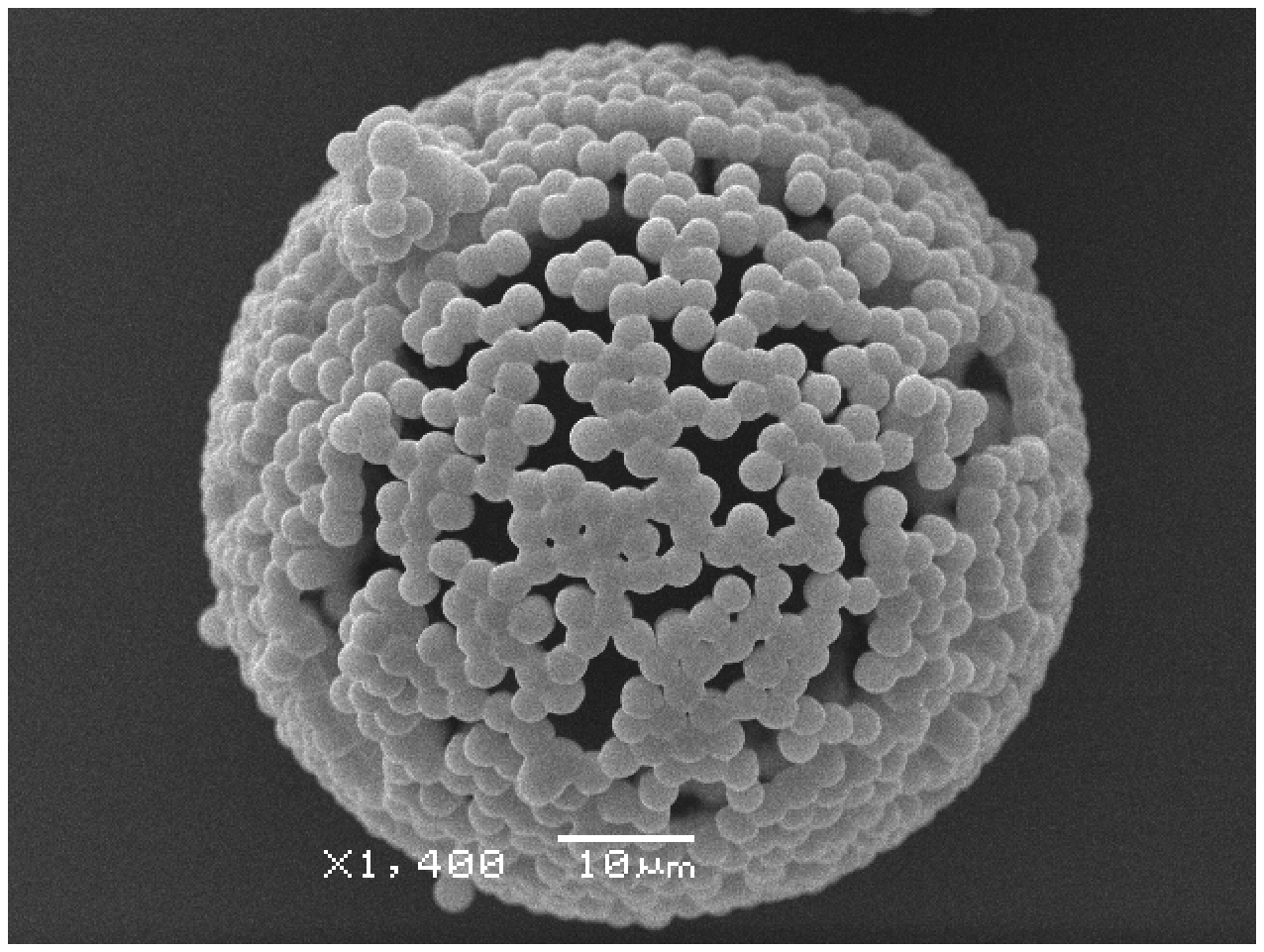}
\includegraphics[width=7.5cm]{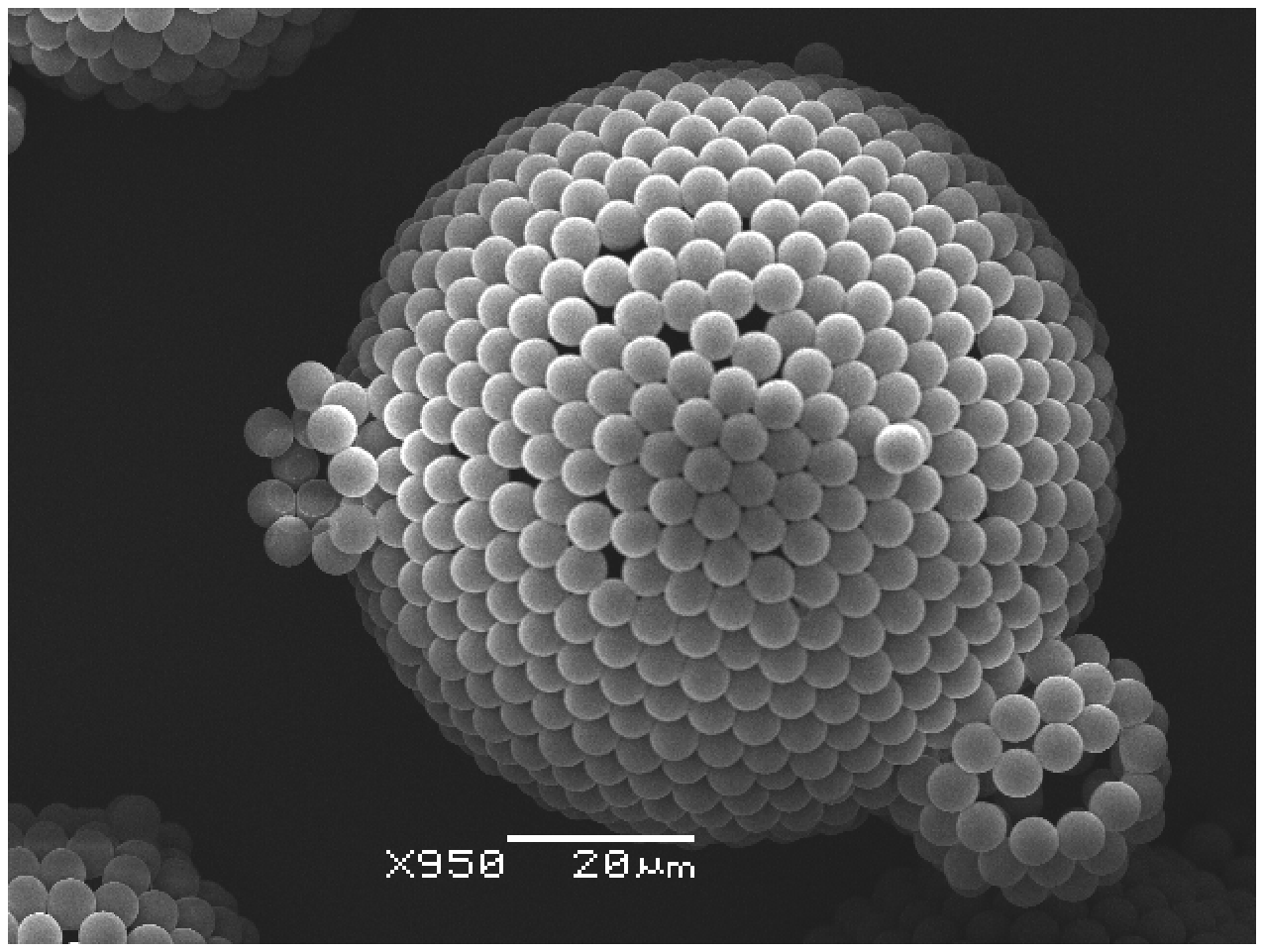}
\end{center}
\caption{SEM images of colloidal cages with a
fluid like particle configuration (left panel, $\rho\sigma^2= 0.84$ and
$D/\sigma = 23.8$) and a crystalline like particle configuration
(right panel, $\rho\sigma^2 = 0.98$ and $D/\sigma = 13.5$).} 
\label{fig:3}
\end{figure}

\begin{figure}[H]
\begin{center}
\includegraphics[width=8cm]{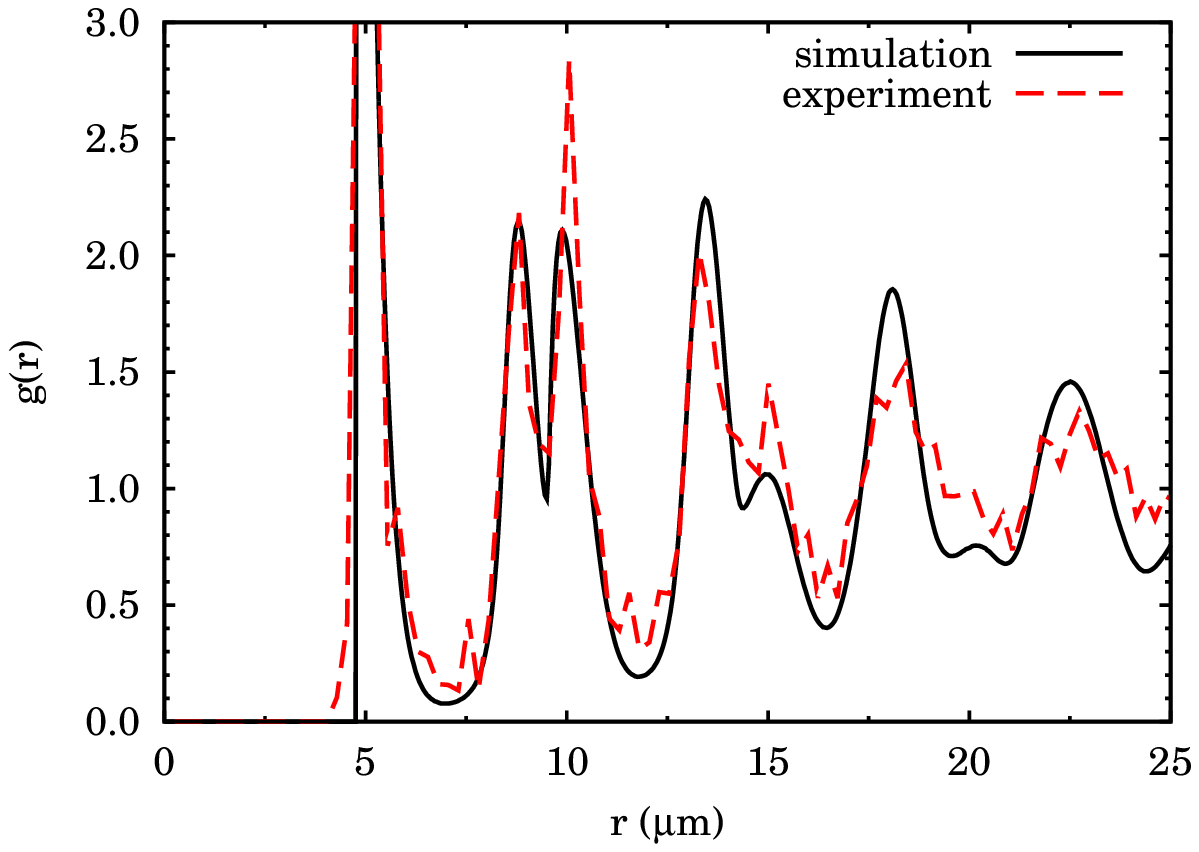}
\includegraphics[width=7cm]{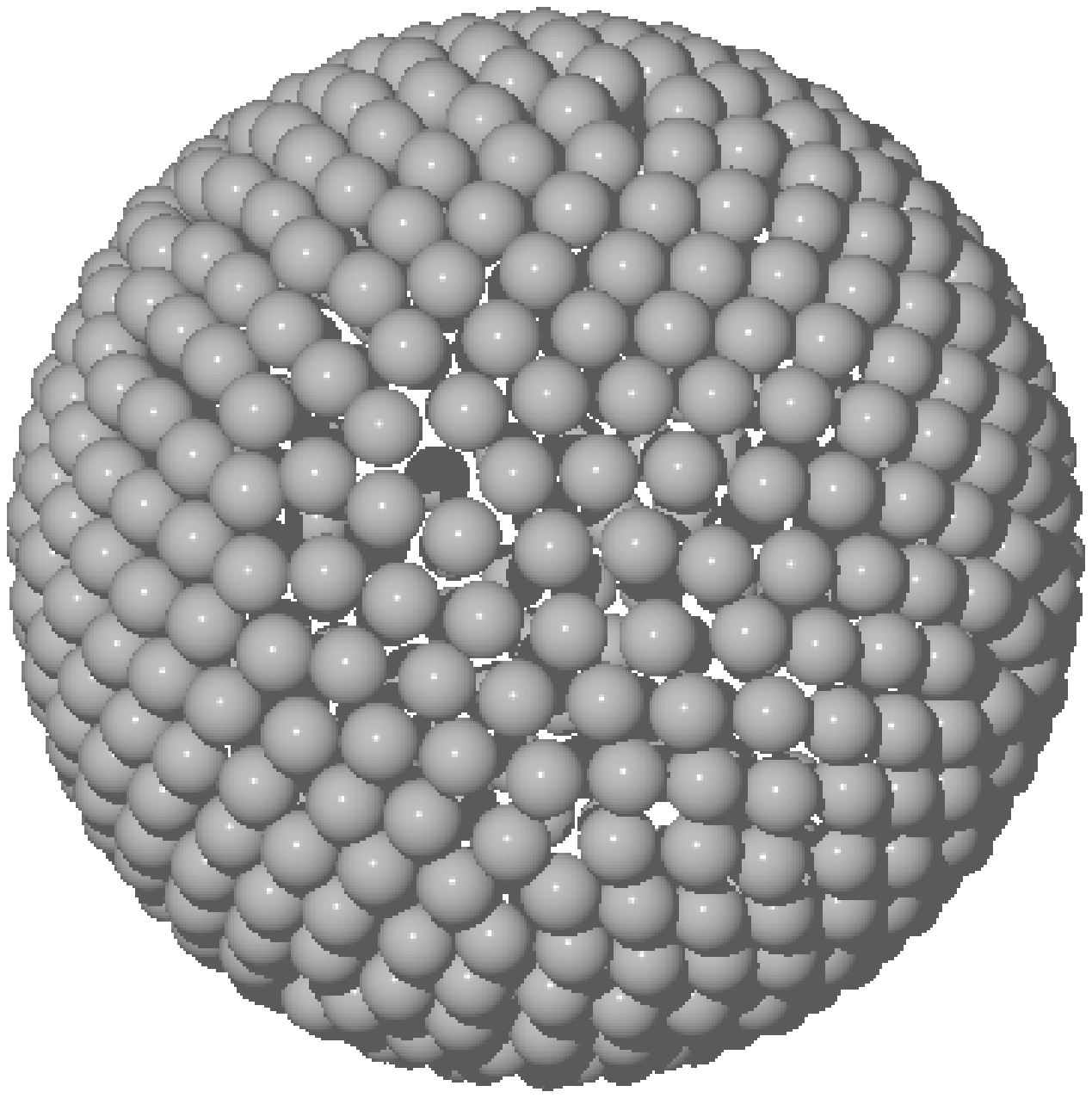}\\
\includegraphics[width=8cm]{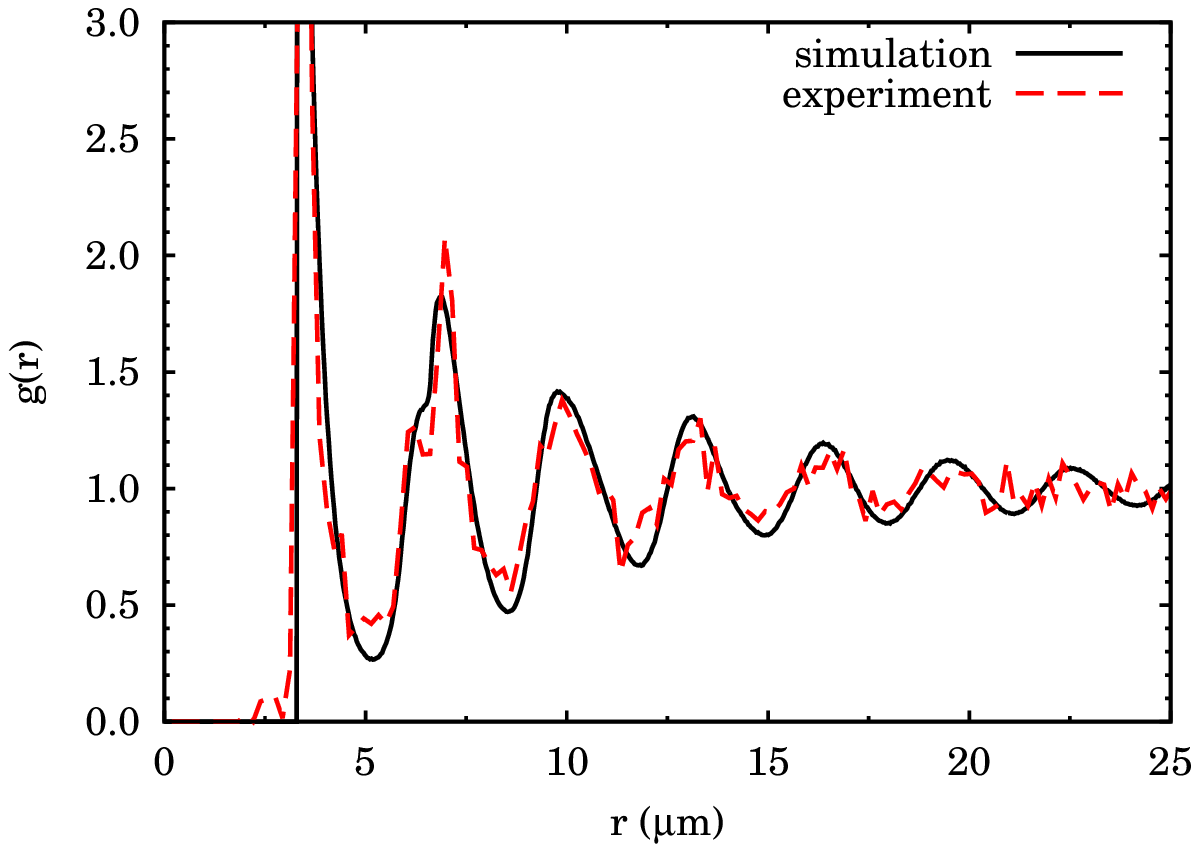}
\includegraphics[width=7cm]{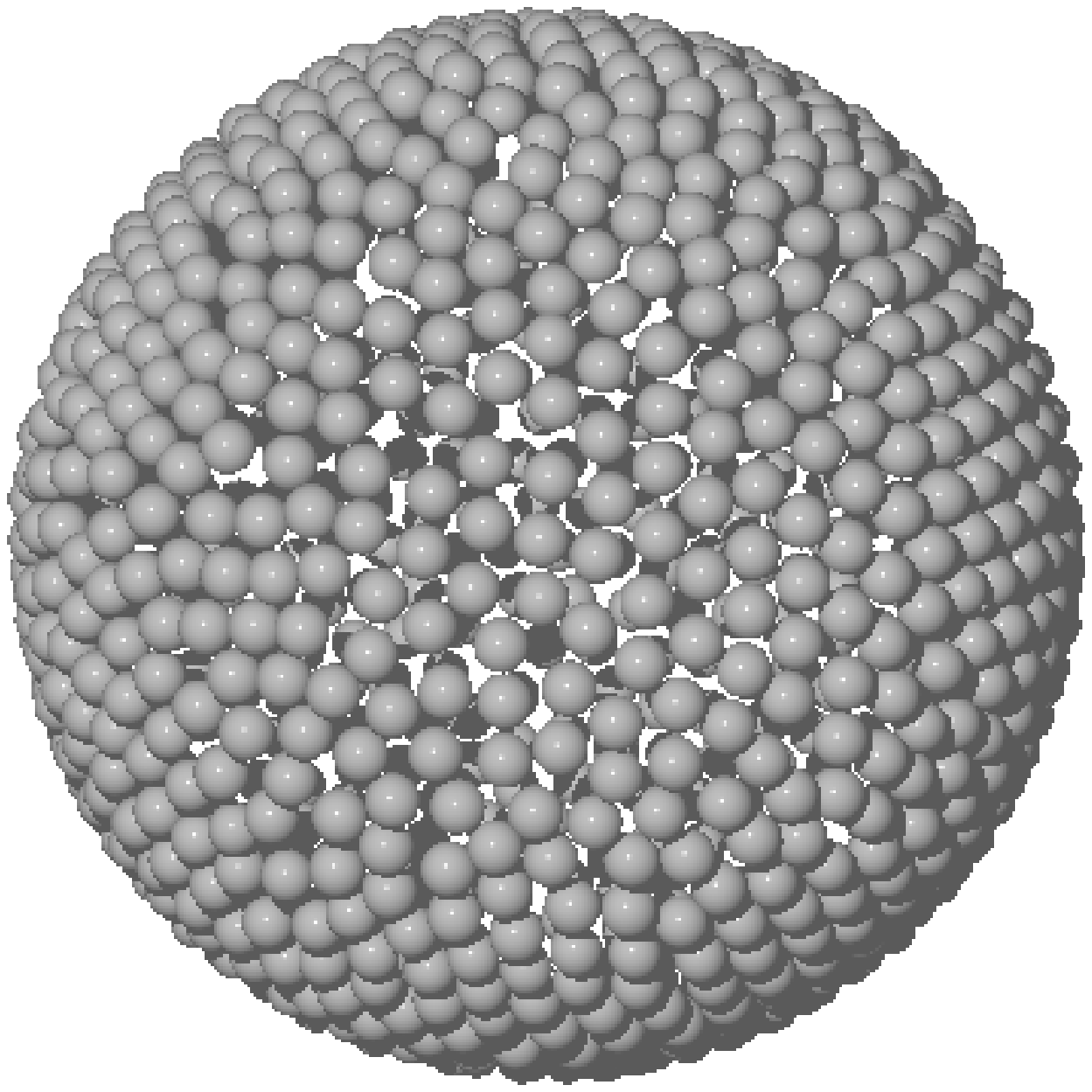}\\
\end{center}
\caption{(Color online) The experimental (dashed red line) and
  simulated (solid black line) RDF of colloidosomes ``a'' (top panel)
  and ``b'' (bottom panel) of Table 
\ref{tab:exp}. The fluid model used in the MC simulations was the
HS. The graphical representation of a snapshot of the particle 
positions when the MC has reached equilibrium, shows \annn{resemblance
with} the SEM images.}  
\label{fig:RDF1}
\end{figure}

\begin{figure}[H]
\begin{center}
\includegraphics[width=8cm]{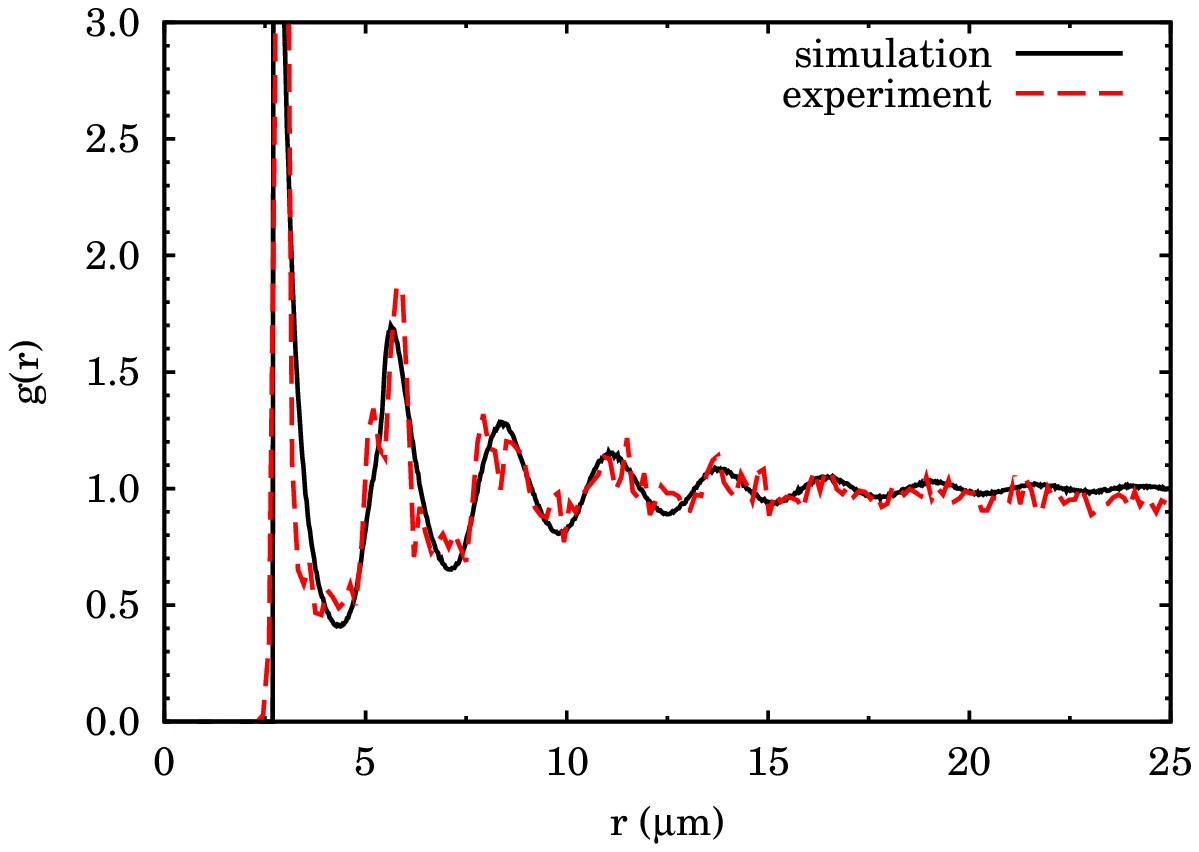}
\includegraphics[width=7cm]{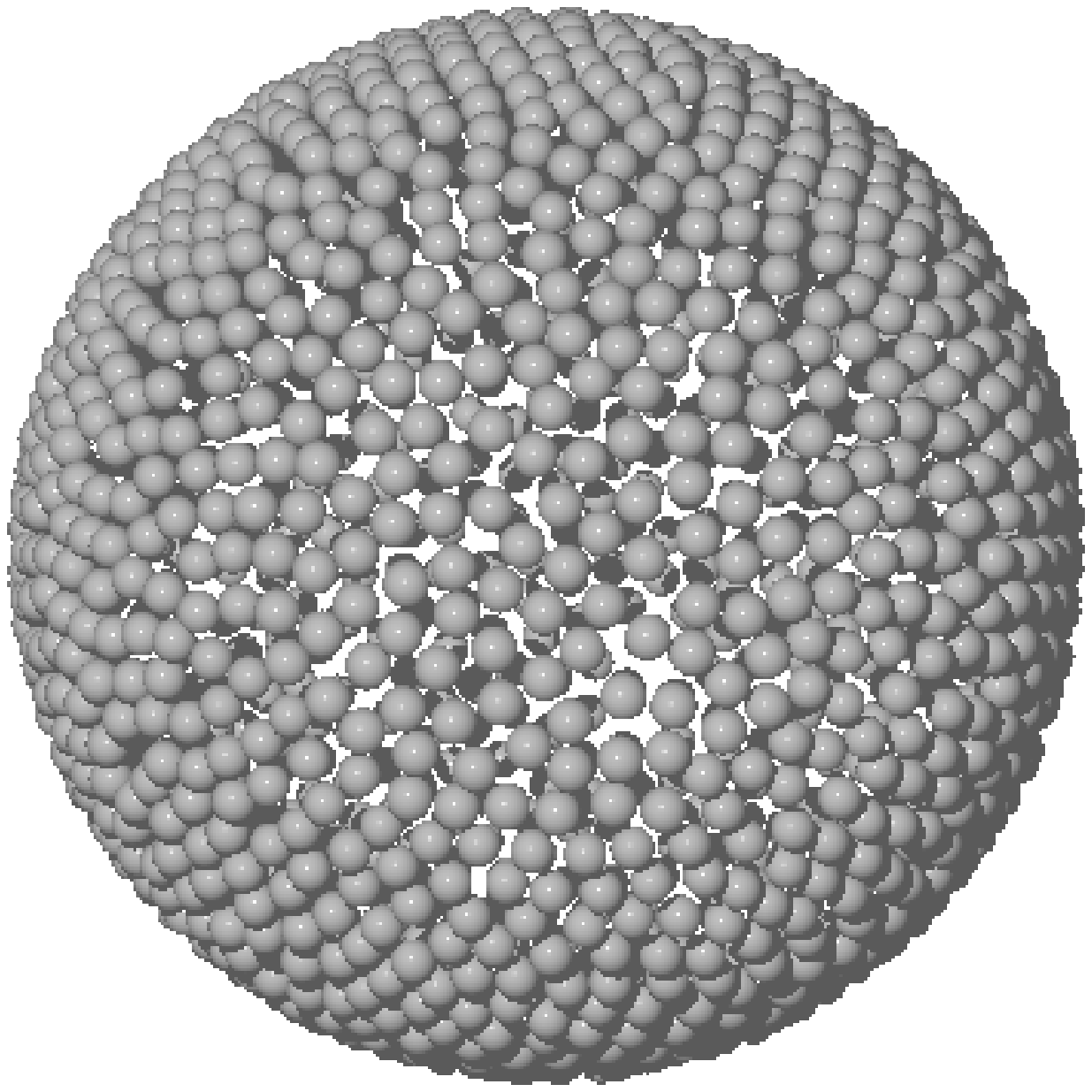}\\
\includegraphics[width=8cm]{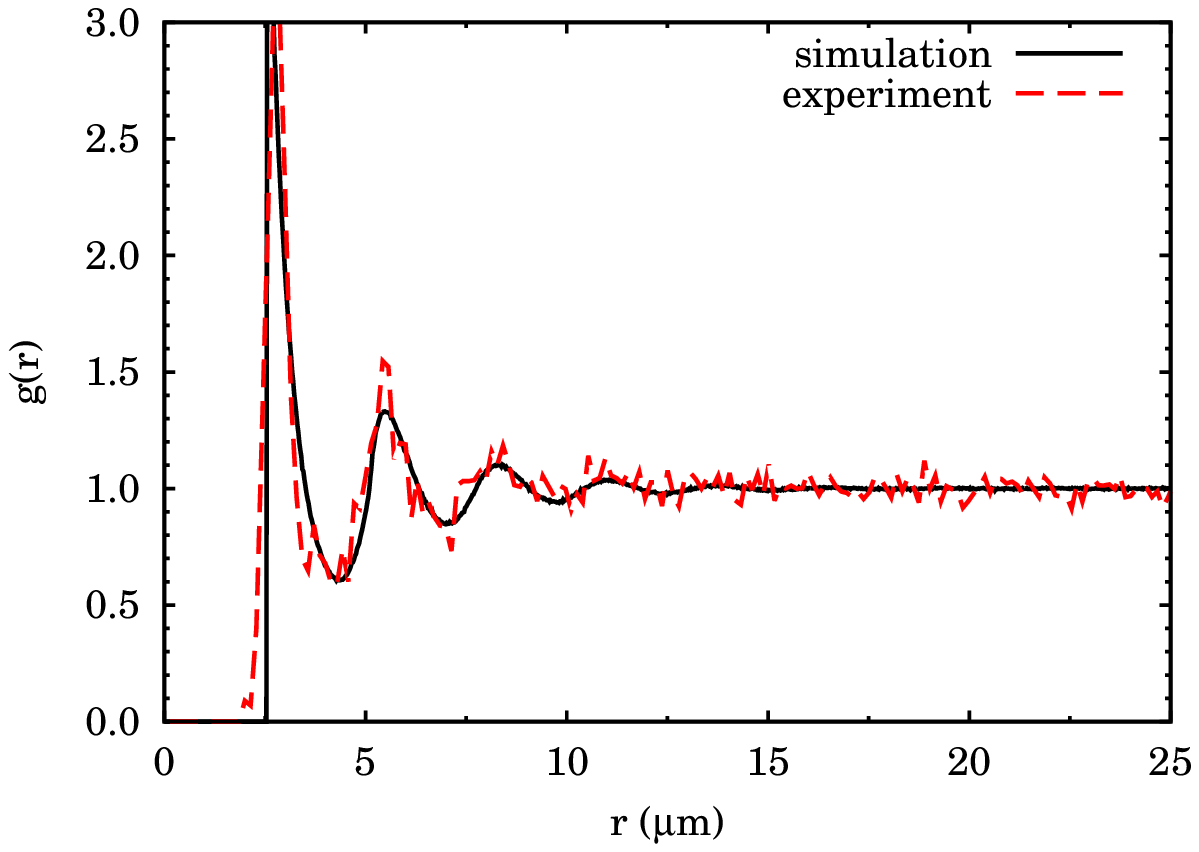}
\includegraphics[width=7cm]{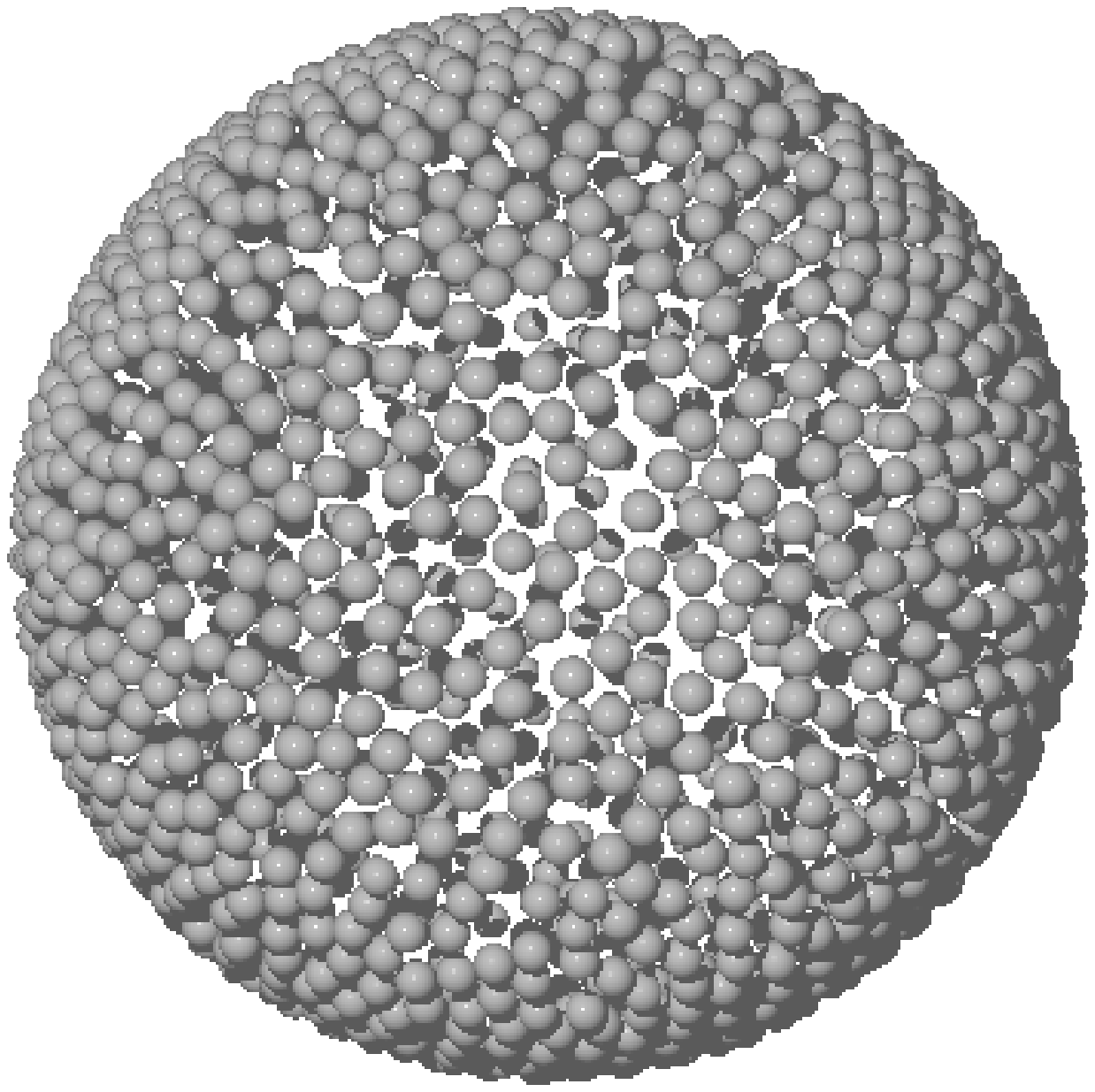}
\end{center}
\caption{(Color online) Same as Fig. \ref{fig:RDF1} for the
  colloidosomes ``c'' (top panel) and ``d'' (bottom panel) of Table
  \ref{tab:exp}.} 
\label{fig:RDF2}
\end{figure}

The results of the experimental colloidosomes are now
compared with the Monte Carlo (MC) simulations. Scanning Electron
Microscopy (SEM) images of two different {colloidal cages} can be found
in Fig. \ref{fig:3}. The RDF of these {colloidal cages} and two others
are shown in Figs. \ref{fig:RDF1} and \ref{fig:RDF2}. The experimental
colloidosomes studied differ one from 
the other by particle size and particle density, the water droplets
were of the same diameter and the temperature was room
temperature, as summarized in Table \ref{tab:exp}. The same 
\annn{values for number of particles, $N$, and sphere diameter,
  $D/\sigma$,} are used in the MC simulations. \annn{Our first choice
  for the pair-potential was the HS fluid model, as justified in
  Section \ref{sec:model}.} 

{It is important to stress that in the experiment we measured the
  RDF from \ann{5 images of different colloidal cages}. Now, there are
  two processes responsible for the assembly of the particles on the
  colloidosome: {\sl i.} the adsorption of the particles on the
  interface at the moment of the formation of the Pickering emulsion
  and {\sl ii.} the motion of the particles on the interface. Our
  experimental measure is clearly not able to discriminate which one
  of the two processes is the more relevant, even if we expect that
  the structure of the colloidal cages obtained after the sintering
  procedure will carry no history of the former process. Moreover
  theoretical studies of process {\sl i.} are, to the best of our
  knowledge \cite{Ehrlich}, much less developed than the ones of the
  latter.} 
\joris{Our computer experiment only takes into account the second
  process assuming that the colloidosome is formed and the
  particles are in thermal equilibrium on the droplet.} 

From Figs. \ref{fig:RDF1} and \ref{fig:RDF2} we can see that there is a
good agreement between the experimental and the theoretical
RDF. This indicates that the HS fluid model gives a good description
of the experimental system. The snapshots from the MC simulation
of the colloidosome differ from the SEM images of the colloidal
cages. The colloidal cages are formed by a network of touching
particles. The structure of the
experimental fluid points to a pretty strong short-range 
attraction between the particles mainly as a result of the sintering
process. The measure of 'structure' used, 
$g(r)$, is not sensitive to these structural differences.

\begin{table}[H]
\begin{ruledtabular}
\begin{tabular}{ccccc}
case &$\sigma$ ($\mu$m) & $N$ & $D/\sigma$ &
$\rho\sigma^2$ \\
\hline
a&4.80	&561	& 13.5 &0.98 \\
b&3.32	&1065	& 19.5 &0.89 \\
c&2.72	&1498	& 23.8 &0.84 \\
d&2.56	&1449	& 25.3 &0.72 \\
\end{tabular}
\end{ruledtabular}
\caption{Characteristics of the experimental colloidosomes
  analyzed. In all cases, the water droplet was of the same diameter
  $D=64.8\,\mu$m. Different colloidosomes differed by the diameter
  $\sigma$ of the colloidal particles and by
  the number $N$ of colloidal particles they carried. The same
  systems, ``a'', ``b'', ``c'', ``d'', have been studied through MC
  simulations.} 
\label{tab:exp}
\end{table}

\begin{figure}[H]
\begin{center}
\includegraphics[width=7.5cm]{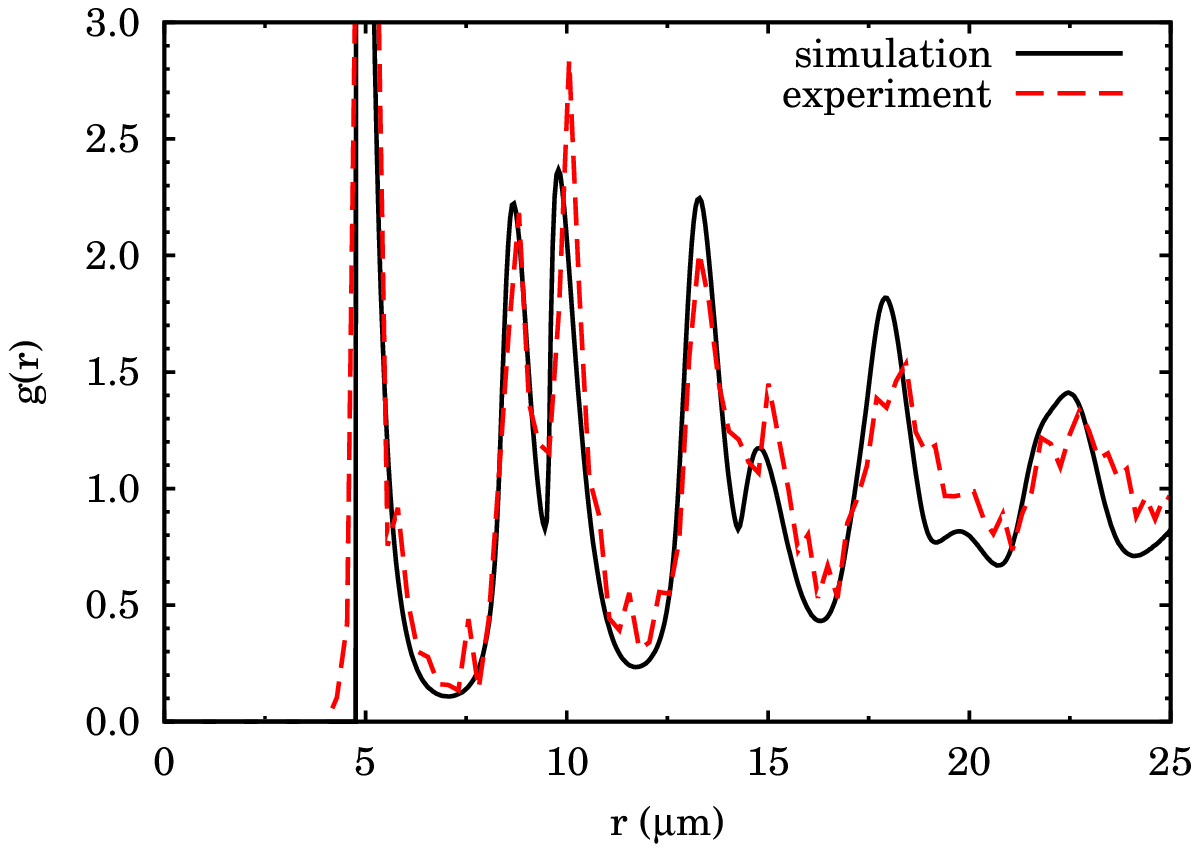}
\includegraphics[width=7.5cm]{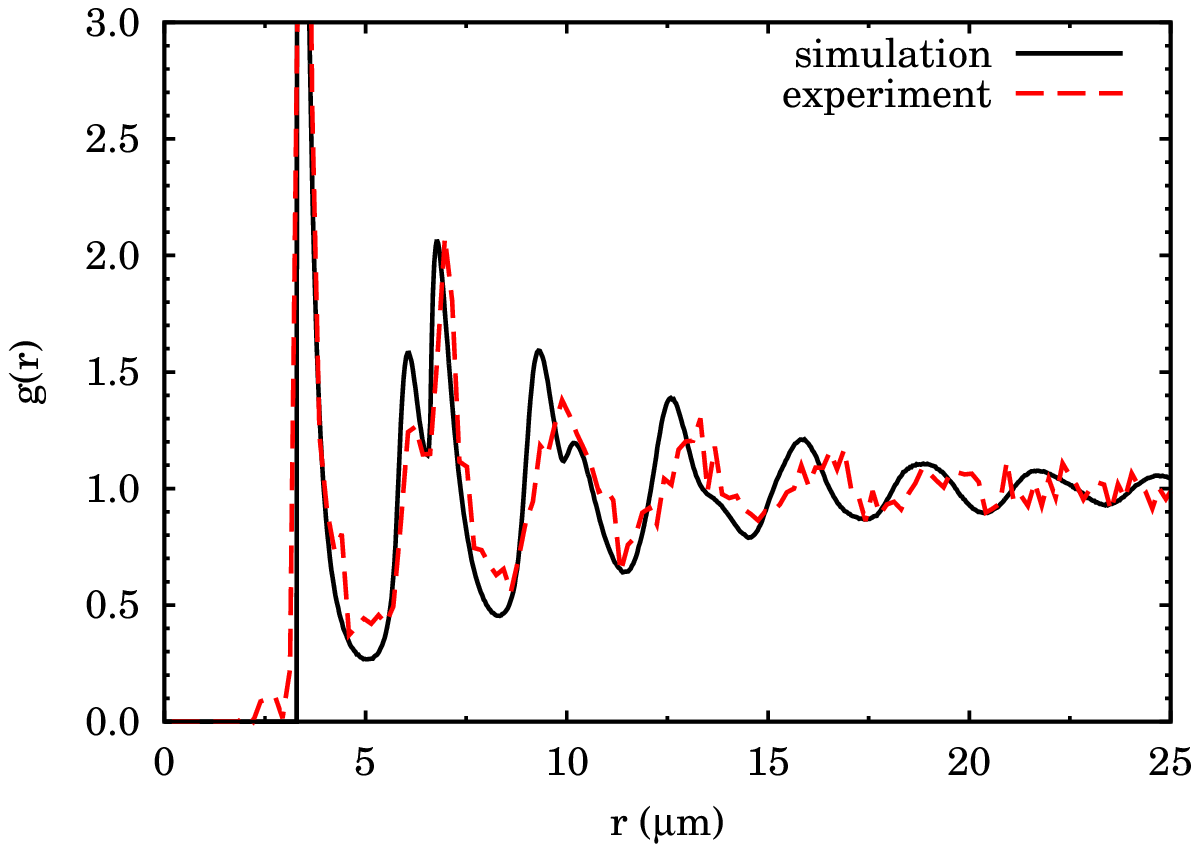}\\
\includegraphics[width=7.5cm]{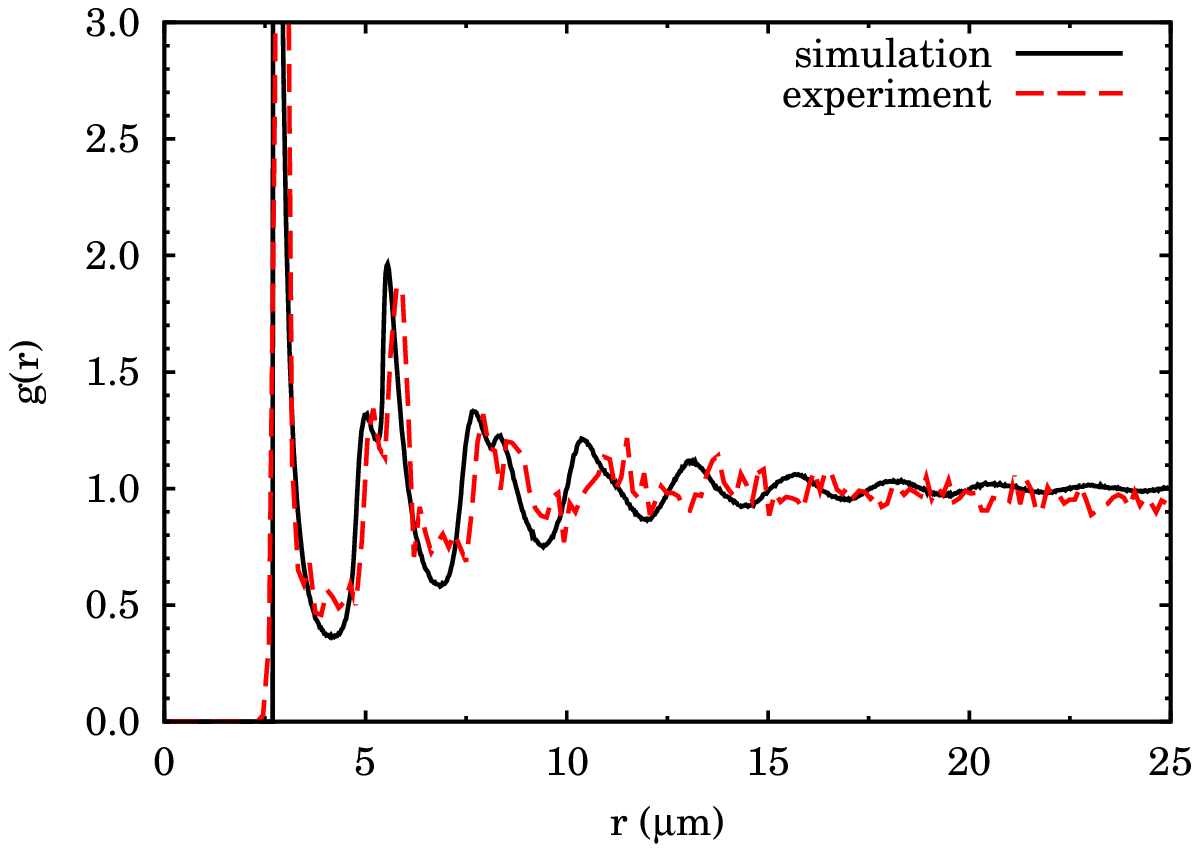}
\includegraphics[width=7.5cm]{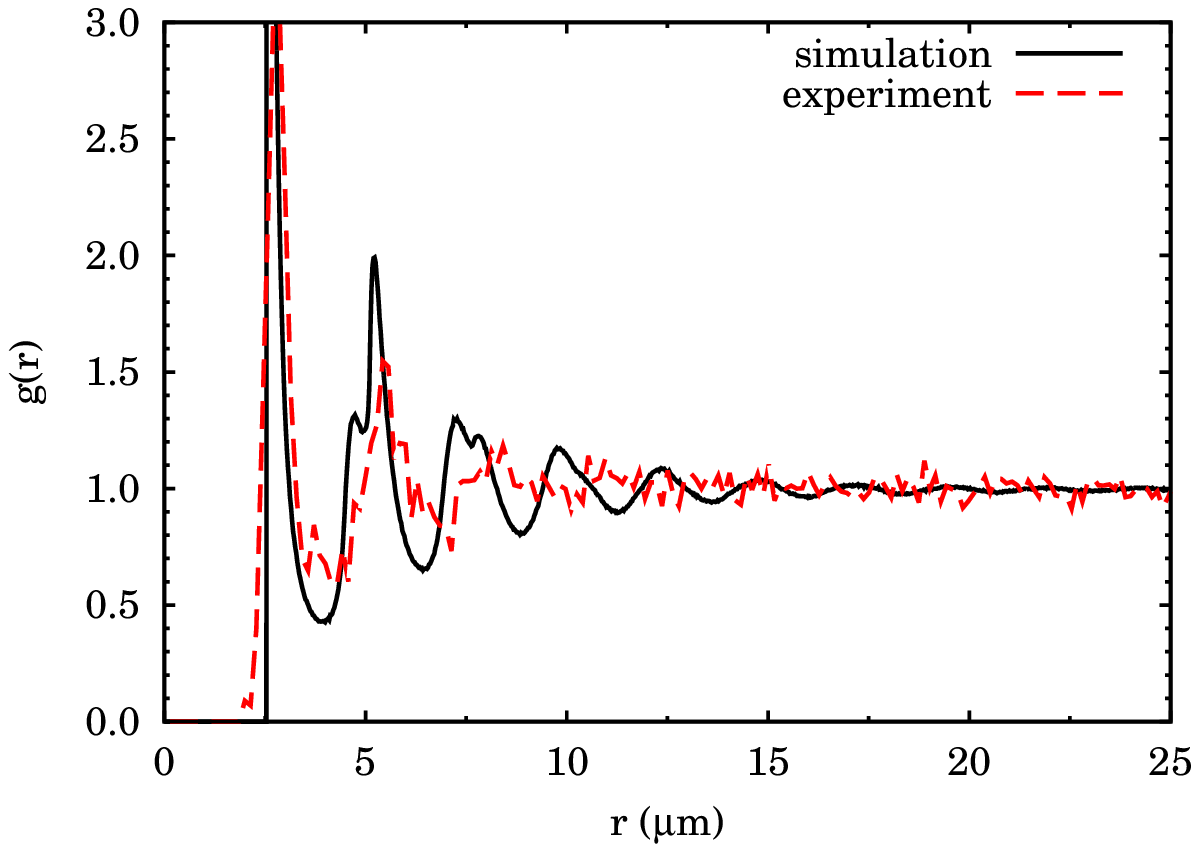}
\end{center}
\caption{(Color online) The experimental (dashed red line) and
  simulated (solid black  
line) RDF of colloidosomes ``a'', ``b'' (top panel) and ``c'', ``d''
(bottom panel) of Table 
\ref{tab:exp}. The fluid model used in the MC simulations was the PHS
with a reduced temperature of $0.3$.}
\label{fig:RDF3}
\end{figure}

We also simulated the experimental colloidosomes with the more
realistic PHS model.     
Initially, the fluid with the highest particle density (``a'') was used to
adjust the reduced temperature. By trial and error we found that
$k_BT/\epsilon_\text{PHS} = 0.3$ 
gave a satisfying agreement with the experiment (see Fig.
\ref{fig:RDF3}). However when we simulated the other colloidal
cages with the same reduced temperature (the experimental temperature
in all cases did not vary and the Hamaker constant did not
change from one colloidosome to the other) we found disagreement
between the MC simulation and the experiment as clearly shown by the
last panel of Fig. \ref{fig:RDF3}. This is an indication that the 
particles used in the experiment do not interact as PHS. An
explanation for this is the balance, in the oil phase, between the
steric repulsion of the polymer chains and the Hamaker attraction. 
\joris{However, steric repulsion through the oil phase is unlikely,
because pVP is insoluble in heptane.
Another, possible explanation is that during emulsification the
attractive interaction is balanced by the shear that is applied, and
this could be reflected on the capsules structures sintered after the
emulsification process.}

In Table \ref{tab:uex} we report the excess internal energy per
particle measured in the MC simulations of the PHS model in
the various systems studied. We can clearly see that as the fluid
develops towards a solid phase there is a lowering of the
energy. 

\begin{table}[H]
\begin{ruledtabular}
\begin{tabular}{cccc}
$k_BT/\epsilon_\text{PHS}$ & $N$ & $D/\sigma$ & $u^{ex}/\epsilon_\text{PHS}$ \\
\hline
0.3&561	 & 13.5 & -2.1509(6)\\
0.3&1065 & 19.5 & -1.862(3)\\
0.3&1498 & 23.8 & -1.732(2)\\
0.3&1449 & 25.3 & -1.586(3)\\
\hline
0.1&1498 & 23.8 & -2.3484(8)\\
1.0&1498 & 23.8 & -1.5450(4)\\
9.1&1498 & 23.8 & -1.5136(3)\\
\hline
9.1&1747 & 23.8 & -2.0932(2)\\
9.1&1248 & 23.8 & -1.0741(2)\\
\end{tabular}
\end{ruledtabular}
\caption{Excess internal energy per particle $u^{ex}=\langle
  U_N\rangle/N$ for the simulated PHS fluids on the sphere.}
\label{tab:uex}
\end{table}
%
%%%%%%%%%%%%%%%%%%%%%%%%%%%%%%%%%%%%%%%%%%%%%%%%%%%%%%%%%%%%%%%%%%%%%%%%%%%%%% 
\section{Conclusions}
%%%%%%%%%%%%%%%%%%%%%%%%%%%%%%%%%%%%%%%%%%%%%%%%%%%%%%%%%%%%%%%%%%%%%%%%%%%%%% 
\label{sec:conclusions}

We have studied a colloidosome of polystyrene hairy particles 
of $\mu$m size moving on the surface of a water droplet in
oil, both experimentally and theoretically through canonical Monte
Carlo simulations, \annn{which is the ensemble of choice for the description
of the experimental system, where the number of particles does not
vary}. In particular we \annn{studied} the radial distribution
function. Agreement 
was found between the experimental \annn{measure} and the
\annn{measure in the computer experiment of the} theoretical 
model of the fluid of pure hard-spheres particles on the surface of a
sphere. We did not find agreement between the experiment and the
polarizable hard-sphere fluid model.  

  It would
  certainly be interesting to pursue a different determination of the
  radial distribution function through the imaging of the same
  colloidosome without going through the sintering procedure. This
  would allow an unbiased determination of the structure of the fluid
  of particles \annn{in their thermal equilibrium} on the droplet
  surface.

Within the \annn{Monte Carlo} simulation, a wide range of particle
densities on the 
colloidosome was studied. At high density, the particles tend to
arrange in a hexagonal lattice, distorted by the curvature of the
droplet and the radial distribution function shows clear signatures of
the first coordination shells. While at low densities a fluid like
behavior is manifested. 

Our \annn{Monte Carlo} simulation results further show that the
addition of an 
attractive tail to the pure hard-sphere pair-potential allows to 
reach solidification by lowering the temperature even at
low densities. We discussed that for pair-potentials with support on
the whole $[0,2R]$, crystallization has to be expected at low
temperatures at any density.   

In our \annn{Monte Carlo} study we have only considered direct
interactions between the 
colloidal particles and not solvent mediated indirect interactions
like excluded volume depletion forces, the Duits interaction between
hairy hard-spheres \cite{Fischer1958}, or the Gourney solvation
interaction, which 
depend on the thermodynamic state of the system \cite{Gazzillo2006}. 
We have simulated two fluid models on the spherical surface: the
athermal HS one and the PHS one. By tuning the reduced temperature in
the PHS model so as to get a structure similar to the one of the
experimental case ``a'' of Table \ref{tab:exp}, we were then unable to
reproduce the experimental radial distribution function of the other
cases ``b'', ``c'', and ``d''. Only the HS model agrees with all four
experimental cases. 

Our simulations shows that the HS fluid model has small correlations
  between particles at opposite poles even at high densities
  \cite{Prestipino1992}, this is not anymore so for particles with a
  soft core \cite{Mendez2008}. From the point of view of our work this
  remains just a theoretical prediction as our experimental measure of
  the radial distribution function is only able to probe half
  hemisphere. Moreover we expect the polystyrene particles used to be
  well described by the hard core pair interaction.
A further interesting
comparison between the experiment and the simulation would be to
compute the orientational correlation function $Q_6$ \cite{Wolde1996}.

A possible further development of the work could be the
realization of the binary mixture of small and large particles on the
water droplet \cite{Law2011} to find experimental evidences for 
demixing predicted by the non additive hard sphere model with negative
non additivity \cite{Santos2010,*Fantoni2011}. Or as a possible way to
push the fluid-like behavior at larger densities, diminishing the
glass gap, as predicted by the additive hard sphere model
\cite{Santos2011}. A natural extension would then be the
multicomponent mixture which in its polydisperse limit may leave no
space to the glass. It would also be possible to simulate the
particles as penetrable-square-well ones \cite{Fantoni09a}.

Colloidosomes may be used to isolate viruses when at the moment of the
formation of the Pickering emulsion only one is captured inside each
droplet. This may be a way to overcome the usual staining
procedure. Of course the opposite situation may also be possible when
many living cells, like for example eukaryotic flagella
\cite{Goldstein2011}, coordinate themselves in the confined geometry of
the drop.

\begin{quotation}
\em
\scriptsize
Simplicio: As for me, I understand in a general way how the two kinds
of natural motions give rise to the circles and spheres; and yet as to
the production of circles by accelerated motion and its proof, I am
not entirely clear; but the fact that one can take the origin of
motion either at the innermost center or at the very top of the sphere
leads one to think that there may be some great mystery hidden in
these true and wonderful results, a mystery related to the creation of
the universe (which is said to be spherical in shape), and related
also to the seat of the first cause.

Salviati: I have no hesitation in agreeing with you. But profound
considerations of this kind belong to a higher science than ours. We
must be satisfied to belong to that class of less worthy workmen who
procure from the quarry the marble out of which, later, the gifted
sculptor produces those masterpieces which are hidden in this rough
and shapeless exterior. Now, if you please, let us proceed.

\centerline{$[\ldots]$}

Sagredo: I really believe that just as, for instance, the few
properties of the circle proven by Euclid in the Third Book of his
Elements lead to many others more recondite, so the principles which
are set forth in this little treatise will, when taken up by
speculative minds, lead to many another more remarkable result; and it
is to be believed that it will be so on account of the nobility of the
subject, which is superior to any other in nature.
\begin{flushright}
\scriptsize
``Dialogues Concerning the Two Sciences'' Galileo Galilei
\end{flushright}
\end{quotation}

\appendix
%%%%%%%%%%%%%%%%%%%%%%%%%%%%%%%%%%%%%%%%%%%%%%%%%%%%%%%%%%%%%%%%%%%%%%%%%%%%%%%
\section{The pair correlation function} 
%%%%%%%%%%%%%%%%%%%%%%%%%%%%%%%%%%%%%%%%%%%%%%%%%%%%%%%%%%%%%%%%%%%%%%%%%%%%%%%
\label{app:1}
Given a classical system of $N$ particles of mass $m$ moving
in a region ${\cal R}$ of a Riemannian manifold of dimension $d$ and
metric tensor $g_{\mu\nu}(\qq)$ with Hamiltonian,
\bq
{\cal H}_N&=&{\cal T}_N+{\cal U}_N~,\\
{\cal T}_N&=&\frac{1}{2m}\sum_{i=1}^Ng^{\mu\nu}(\qq_i)p_{i\mu}p_{i\nu}~,\\  
{\cal U}_N&=&\sum U_N(\qq_1,\ldots,\qq_N)~,
\eq
where we denote with $\qq=(q^1,\ldots,q^d)$ a point
of the manifold, with $\qq_i=(q^1_i,\ldots,q^d_i)$ the coordinates of
particle $i$ and with $\pp_i=(p_{i1},\ldots,p_{id})$ its canonically
conjugate momenta, and we use the Einstein summation convention
to omit the sum over the repeated Greek indices.
The canonical ensemble probability density to find the statistical
system of distinguishable particles, the {\sl fluid}, in
thermal equilibrium at an inverse temperature $\beta=1/k_BT$ (with
$k_B$ Boltzmann constant) with coordinates 
$\QQ=(\qq_1,\ldots,\qq_N)$ and momenta
$\PP=(\pp_1,\ldots,\pp_N)$ is 
\bq
{\cal F}(\QQ,\PP,N)=\frac{1}{\Theta}\frac{1}{h^{dN}N!}
e^{-\beta({\cal T}_N+{\cal U}_N)}~,
\eq
where $h$ is Planck constant and the normalization factor $\Theta$ is
the partition function of the canonical ensemble of the identical
particles 
\bq
\Theta=\frac{1}{h^{dN}N!}\int_{{\cal R}^N}d\QQ\int d\PP 
e^{-\beta({\cal T}_N+{\cal U}_N)}=
\frac{1}{\Lambda^{dN}N!}\int_{{\cal R}^N}d\VV e^{-\beta {\cal U}_N}=e^{-\beta F}~,
\eq
where in the second equality we performed the Gaussian integral over
the conjugated momenta so that $d\VV=d\vv_1\cdots d\vv_N$ with
$d\vv_i=\sqrt{g}\prod_{\mu=1}^ddq_i^\mu$
(here $g=\det||g_{\mu\nu}(\qq_i)||=[\det||g^{\mu\nu}(\qq_i)||]^{-1}$)
the infinitesimal ``volume'' element of the manifold and
$\Lambda=\sqrt{2\pi\beta\hbar^2/m}$ is the de Broglie thermal
wavelength. To justify a classical treatment of the statistical
properties it is necessary that $\Lambda$ be much less than the mean
nearest neighbor distance between the particles. In the last equality
we used the definition of entropy and $F$ is the Helmholtz free energy. 

If the particles move on a sphere of radius one, ${\cal R}=S_1$, then
the coordinates are the polar 
coordinates on the sphere $\qq_i=(\theta_i,\varphi_i)$ with
$\theta_i\in [0,\pi]$ the polar angle and $\varphi_i\in [0,2\pi)$ the
azimuthal angle, and the metric tensor is
\bq
||g_{\mu\nu}(\qq)||=\left(
\begin{array}{cc}
1 & 0 \\
0 & \sin^2\theta
\end{array}
\right)~,
\eq
so that $\det||g_{\mu\nu}(\qq_i)||=\sin^2\theta_i$ and
$d\vv_i=\sin\theta_id\theta_id\varphi_i\equiv d\Omega_i$ the solid
angle spanned by the three dimensional vector $\rr_i$ of the position
of particle $i$ in the Euclidean space with origin on the sphere
center. Given a physical observable  
only function of the coordinates ${\cal O}_N=O_N(\QQ)$, we can then
measure its average value as
\bq
\langle{\cal O}_N\rangle=\frac{\int_{{\cal R}^N} O_N(\QQ)
e^{-\beta U_N(\QQ)}\,\prod_{i=1}^Nd\vv_i}
{\int_{{\cal R}^N} e^{-\beta U_N(\QQ)}\,\prod_{i=1}^Nd\vv_i}~.
\eq
 
For example the one body correlation function \cite{Hill} for the
particles on a sphere of radius $R$ is measured as
\bq \label{rho1}
\rho(\qq)=\left\langle\sum_{i=1}^N\frac{\delta^{(2)}(\qq,\qq_i)}{R^2}\right
\rangle~,
\eq
where
$\delta^{(2)}(\qq,\qq^\prime)=\delta(\theta-\theta^\prime)
\delta(\varphi-\varphi^\prime)/\sqrt{g}$ 
is the Dirac delta function on the manifold. We now use the fact that
our potential energy (\ref{tpot}) is invariant under any rotation of the
reference frame to say that $\rho(\qq)$ has to
be independent of $\qq$ and (by integrating (\ref{rho1}) over $d\vv$)
we must have $\rho(\qq)=\rho=N/(4\pi R^2)$.  

The two body correlation function \cite{Hill} is measured as
\bq \label{rho2}
g(\qq,\qq^\prime)=\left\langle\sum_{i\neq j}
\frac{\delta^{(2)}(\qq,\qq_i)}{R^2}
\frac{\delta^{(2)}(\qq^\prime,\qq_j)}{R^2}
\right\rangle\Big/[\rho(\qq)\rho(\qq^\prime)]~.
\eq
Because of rotational invariance $g$ can only depend on the geodesic
distance $d$ between $\qq$ and $\qq^\prime$. We can then calculate it
on a reference frame where $\varphi=\varphi^\prime$ so that
$d=R(\theta-\theta^\prime)$ and
\bq \label{rho2i}
g(d)=\left\langle\sum_{i\neq j}
\frac{\delta(\theta-\theta_{i})}{R^2\sin\theta_i}
\frac{\delta(\theta^\prime-\theta_{j})}{\sin\theta^\prime}
\delta(\varphi^\prime-\varphi_i)\delta(\varphi^\prime-\varphi_j)
\right\rangle\Big/(R^2\rho^2)~.
\eq 
If we now choose $\theta=\bar{\theta}+\theta^\prime$ and integrate
over $d\vv^\prime$ we get
\bq
g(d)=\left\langle\sum_{i\neq j}
\frac{\delta(\bar{\theta}-\theta_{ij})}{R^2\sin\theta_i}
\delta(\varphi_{ij})
\right\rangle\Big/(N\rho)~,
\eq 
where $\theta_{ij}=\theta_i-\theta_j$ and $\varphi_{ij}=\varphi_i-\varphi_j$.
We can use rotational invariance to choose the sphere north pole
sitting on particle $j$ to get further
\bq
g(d)=\left\langle\sum_{i\neq j}
\delta(\bar{\theta}-\theta_{ij})\delta(\varphi_{ij})
\right\rangle\Big/(N\rho R^2\sin\bar{\theta})~.
\eq
In place of the geodesic distance $R\bar{\theta}$ we can use the
Euclidean distance $d=2R\sin(\bar{\theta}/2)$. We can then use the equality
$\delta(\bar{\theta}-\theta_{ij})=\delta(d-d_{ij})R\cos(\bar{\theta}/2)$, here
$d_{ij}=2R\sin(\theta_{ij}/2)$, to write 
\bq \nonumber
g(d)&=&\left\langle\sum_{i\neq j}
\delta(d-d_{ij})\delta(\varphi_{ij})
\right\rangle\Big/(N\rho d)\\
&=&\left\langle\sum_{i\neq j}
\delta(d-r_{ij})\delta(\varphi_{ij})
\right\rangle\Big/(N\rho d)~,
\eq
where $r_{ij}$ is defined in Eq. (\ref{ed}) of the main text.
Now we use rotational invariance noticing again that given any two
point $\qq_i$ and $\qq_j$ on the sphere one can always find a
reference frame in which $\varphi_i=\varphi_j$ to get
\bq \label{gr}
g(d)=\left\langle\sum_{i\neq j}
\delta(d-r_{ij})
\right\rangle\Big/(N\rho 2\pi d)~.
\eq

%%%%%%%%%%%%%%%%%%%%%%%%%%%%%%%%%%%%%%%%%%%%%%%%%%%%%%%%%%%%%%%%%%%%%%%%%%%%%%%
\section{The Monte Carlo simulation} 
%%%%%%%%%%%%%%%%%%%%%%%%%%%%%%%%%%%%%%%%%%%%%%%%%%%%%%%%%%%%%%%%%%%%%%%%%%%%%%%
\label{app:2}
%%%%%%%%%%%%%%%%%%%%%%%%%%%%%%%%%%%%%%%%%%%%%%%%%%%%%%%%%%%%%%%%%%%%%%%%%%%%%% 
In the Monte Carlo integration one does a random walk
\cite{Kalos-Whitlock} in $\QQ$ with 
$\theta_i\in[0,\pi], \varphi_i\in[-\pi/2,\pi/2)$ for all
$i=1,\ldots,N$ with periodic boundary conditions:
$\varphi=\varphi+2\pi$ and $\theta=\theta+\pi$. In the Metropolis
algorithm \cite{Kalos-Whitlock} one takes as the acceptance probability
\bq
A[\QQ\to \QQ^\prime]=
\min\left\{1,e^{-\beta[U_N(\QQ^\prime)-U_N(\QQ)]}
\frac{\prod_{i=1}^N\sin\theta_i^\prime}
{\prod_{i=1}^N\sin \theta_i}\right\}~.
\eq

\begin{acknowledgments}
The MC simulations were carried out at the Center for High Performance
Computing (CHPC), CSIR Campus, 15 Lower Hope St., Rosebank, Cape Town,
South Africa. 
B. K. acknowledges support by the South African Research Chairs
Initiative of the Department of Science and Technology and National
Research Foundation. 
\end{acknowledgments}
%%%%%%%%%%%%%%%%%%%%%%%%%%%%%%%%%%%%%%%%%%%%%%%%%%%%%%%%%%%%%%%%%%%%%%%%%%%%%%%
%\bibliographystyle{apsrev}
%\bibliography{somes}

\begin{thebibliography}{64}%
\makeatletter
\providecommand \@ifxundefined [1]{%
 \@ifx{#1\undefined}
}%
\providecommand \@ifnum [1]{%
 \ifnum #1\expandafter \@firstoftwo
 \else \expandafter \@secondoftwo
 \fi
}%
\providecommand \@ifx [1]{%
 \ifx #1\expandafter \@firstoftwo
 \else \expandafter \@secondoftwo
 \fi
}%
\providecommand \natexlab [1]{#1}%
\providecommand \enquote  [1]{``#1''}%
\providecommand \bibnamefont  [1]{#1}%
\providecommand \bibfnamefont [1]{#1}%
\providecommand \citenamefont [1]{#1}%
\providecommand \href@noop [0]{\@secondoftwo}%
\providecommand \href [0]{\begingroup \@sanitize@url \@href}%
\providecommand \@href[1]{\@@startlink{#1}\@@href}%
\providecommand \@@href[1]{\endgroup#1\@@endlink}%
\providecommand \@sanitize@url [0]{\catcode `\\12\catcode `\$12\catcode
  `\&12\catcode `\#12\catcode `\^12\catcode `\_12\catcode `\%12\relax}%
\providecommand \@@startlink[1]{}%
\providecommand \@@endlink[0]{}%
\providecommand \url  [0]{\begingroup\@sanitize@url \@url }%
\providecommand \@url [1]{\endgroup\@href {#1}{\urlprefix }}%
\providecommand \urlprefix  [0]{URL }%
\providecommand \Eprint [0]{\href }%
\providecommand \doibase [0]{http://dx.doi.org/}%
\providecommand \selectlanguage [0]{\@gobble}%
\providecommand \bibinfo  [0]{\@secondoftwo}%
\providecommand \bibfield  [0]{\@secondoftwo}%
\providecommand \translation [1]{[#1]}%
\providecommand \BibitemOpen [0]{}%
\providecommand \bibitemStop [0]{}%
\providecommand \bibitemNoStop [0]{.\EOS\space}%
\providecommand \EOS [0]{\spacefactor3000\relax}%
\providecommand \BibitemShut  [1]{\csname bibitem#1\endcsname}%
\let\auto@bib@innerbib\@empty
%</preamble>
\bibitem [{\citenamefont {{A. D. Dinsmore {\sl et al.}}}(2002)}]{Dinsmore2002}%
  \BibitemOpen
  \bibfield  {author} {\bibinfo {author} {\bibnamefont {{A. D. Dinsmore {\sl et
  al.}}}},\ }\href@noop {} {\bibfield  {journal} {\bibinfo  {journal}
  {Science}\ }\textbf {\bibinfo {volume} {298}},\ \bibinfo {pages} {1006}
  (\bibinfo {year} {2002})}\BibitemShut {NoStop}%
\bibitem [{\citenamefont {{C. Zeng, H. Bissig, and A. D.
  Dinsmore}}(2006)}]{Zeng2006}%
  \BibitemOpen
  \bibfield  {author} {\bibinfo {author} {\bibnamefont {{C. Zeng, H. Bissig,
  and A. D. Dinsmore}}},\ }\href@noop {} {\bibfield  {journal} {\bibinfo
  {journal} {Solid State Comm.}\ }\textbf {\bibinfo {volume} {139}},\ \bibinfo
  {pages} {547} (\bibinfo {year} {2006})}\BibitemShut {NoStop}%
\bibitem [{\citenamefont {{A. B. Subramanian, M. Abkarian, and H. A. Stone
  }}(2005)}]{Subramanian2005}%
  \BibitemOpen
  \bibfield  {author} {\bibinfo {author} {\bibnamefont {{A. B. Subramanian, M.
  Abkarian, and H. A. Stone }}},\ }\href@noop {} {\bibfield  {journal}
  {\bibinfo  {journal} {Nature Mat. Lett.}\ }\textbf {\bibinfo {volume} {4}},\
  \bibinfo {pages} {553} (\bibinfo {year} {2005})}\BibitemShut {NoStop}%
\bibitem [{\citenamefont {{R. F. Lee}}(1999)}]{Lee1999}%
  \BibitemOpen
  \bibfield  {author} {\bibinfo {author} {\bibnamefont {{R. F. Lee}}},\
  }\href@noop {} {\bibfield  {journal} {\bibinfo  {journal} {Spill Science and
  Technology Bulletin}\ }\textbf {\bibinfo {volume} {5}},\ \bibinfo {pages}
  {117} (\bibinfo {year} {1999})}\BibitemShut {NoStop}%
\bibitem [{\citenamefont {{E. Dickinson}}(2010)}]{Dickinson2010}%
  \BibitemOpen
  \bibfield  {author} {\bibinfo {author} {\bibnamefont {{E. Dickinson}}},\
  }\href@noop {} {\bibfield  {journal} {\bibinfo  {journal} {Current Opinion in
  Colloid and Interface Science}\ }\textbf {\bibinfo {volume} {15}},\ \bibinfo
  {pages} {40} (\bibinfo {year} {2010})}\BibitemShut {NoStop}%
\bibitem [{\citenamefont {{D. Rousseau, S. Ghosh, and H.
  Park}}(2009)}]{Rousseau2009}%
  \BibitemOpen
  \bibfield  {author} {\bibinfo {author} {\bibnamefont {{D. Rousseau, S. Ghosh,
  and H. Park}}},\ }\href@noop {} {\bibfield  {journal} {\bibinfo  {journal}
  {Journal of Food Science}\ }\textbf {\bibinfo {volume} {74}},\ \bibinfo
  {pages} {E1} (\bibinfo {year} {2009})}\BibitemShut {NoStop}%
\bibitem [{\citenamefont {{S. Lu, R. J. Pugh, and E.
  Forssberg}}(2005)}]{Lu2005}%
  \BibitemOpen
  \bibfield  {author} {\bibinfo {author} {\bibnamefont {{S. Lu, R. J. Pugh, and
  E. Forssberg}}},\ }in\ \href@noop {} {\emph {\bibinfo {booktitle}
  {Interfacial separation of particles}}},\ Vol.~\bibinfo {volume} {20}\
  (\bibinfo  {publisher} {Elsevier},\ \bibinfo {address} {Amsterdam},\ \bibinfo
  {year} {2005})\BibitemShut {NoStop}%
\bibitem [{\citenamefont {{J. Frelichowska, M. A. Bolzinger, J. Pelletier, J.
  P. Valour, and Y. Chevalier}}(2009)}]{Frelichowska2009}%
  \BibitemOpen
  \bibfield  {author} {\bibinfo {author} {\bibnamefont {{J. Frelichowska, M. A.
  Bolzinger, J. Pelletier, J. P. Valour, and Y. Chevalier}}},\ }\href@noop {}
  {\bibfield  {journal} {\bibinfo  {journal} {International Journal of
  Pharmaceutics}\ }\textbf {\bibinfo {volume} {371}},\ \bibinfo {pages} {56}
  (\bibinfo {year} {2009})}\BibitemShut {NoStop}%
\bibitem [{\citenamefont {{P. H. F. Hansen, S. R\"odner, and L.
  Bergstr\"om}}(2001)}]{Hansen2001}%
  \BibitemOpen
  \bibfield  {author} {\bibinfo {author} {\bibnamefont {{P. H. F. Hansen, S.
  R\"odner, and L. Bergstr\"om}}},\ }\href@noop {} {\bibfield  {journal}
  {\bibinfo  {journal} {Langmuir}\ }\textbf {\bibinfo {volume} {17}},\ \bibinfo
  {pages} {4867} (\bibinfo {year} {2001})}\BibitemShut {NoStop}%
\bibitem [{\citenamefont {{P. Binks and T. S. Horozov}}(2006)}]{Binks2006}%
  \BibitemOpen
  \bibfield  {author} {\bibinfo {author} {\bibnamefont {{P. Binks and T. S.
  Horozov}}},\ }\href@noop {} {\emph {\bibinfo {title} {Colloidal particles at
  liquid interfaces}}}\ (\bibinfo  {publisher} {Cambridge University Press},\
  \bibinfo {address} {New York},\ \bibinfo {year} {2006})\BibitemShut {NoStop}%
\bibitem [{\citenamefont {{L. Hong, S. Jiang, and S.
  Granick}}(2006)}]{Hong2006}%
  \BibitemOpen
  \bibfield  {author} {\bibinfo {author} {\bibnamefont {{L. Hong, S. Jiang, and
  S. Granick}}},\ }\href@noop {} {\bibfield  {journal} {\bibinfo  {journal}
  {Langmuir}\ }\textbf {\bibinfo {volume} {22}},\ \bibinfo {pages} {9495}
  (\bibinfo {year} {2006})}\BibitemShut {NoStop}%
\bibitem [{\citenamefont {{S. A. F. Bon and T. Chen}}(2007)}]{Bon2007}%
  \BibitemOpen
  \bibfield  {author} {\bibinfo {author} {\bibnamefont {{S. A. F. Bon and T.
  Chen}}},\ }\href@noop {} {\bibfield  {journal} {\bibinfo  {journal}
  {Langmuir}\ }\textbf {\bibinfo {volume} {23}},\ \bibinfo {pages} {9527}
  (\bibinfo {year} {2007})}\BibitemShut {NoStop}%
\bibitem [{\citenamefont {{E. Pitard, M. L. Rosinberg, G. Stell, and G.
  Tarjus}}(1995)}]{Pitard1995}%
  \BibitemOpen
  \bibfield  {author} {\bibinfo {author} {\bibnamefont {{E. Pitard, M. L.
  Rosinberg, G. Stell, and G. Tarjus}}},\ }\href@noop {} {\bibfield  {journal}
  {\bibinfo  {journal} {Phys. Rev. Lett.}\ }\textbf {\bibinfo {volume} {74}},\
  \bibinfo {pages} {4361} (\bibinfo {year} {1995})}\BibitemShut {NoStop}%
\bibitem [{\citenamefont {{Isabelle Cantat, Sylvie Cohen-Addad, Florence Elias,
  Fran\c{c}ois Graner, Renihard H\"ohler, Olivier Pitois, Florence Rouyer, and
  Arnaud Saint-Jalmes}}(2010)}]{Choen-Addad}%
  \BibitemOpen
  \bibfield  {author} {\bibinfo {author} {\bibnamefont {{Isabelle Cantat,
  Sylvie Cohen-Addad, Florence Elias, Fran\c{c}ois Graner, Renihard H\"ohler,
  Olivier Pitois, Florence Rouyer, and Arnaud Saint-Jalmes}}},\ }\href@noop {}
  {\emph {\bibinfo {title} {Les mousses: structure et dynamique}}}\ (\bibinfo
  {publisher} {Belin},\ \bibinfo {year} {2010})\BibitemShut {NoStop}%
\bibitem [{\citenamefont {{P. Pieranski}}(1980)}]{Pieranski1980}%
  \BibitemOpen
  \bibfield  {author} {\bibinfo {author} {\bibnamefont {{P. Pieranski}}},\
  }\href@noop {} {\bibfield  {journal} {\bibinfo  {journal} {Phys. Rev. Lett.}\
  }\textbf {\bibinfo {volume} {45}},\ \bibinfo {pages} {569} (\bibinfo {year}
  {1980})}\BibitemShut {NoStop}%
\bibitem [{\citenamefont {{R. Aveyard {\sl et al.}}}(2002)}]{Aveyard2002}%
  \BibitemOpen
  \bibfield  {author} {\bibinfo {author} {\bibnamefont {{R. Aveyard {\sl et
  al.}}}},\ }\href@noop {} {\bibfield  {journal} {\bibinfo  {journal} {Phys.
  Rev. Lett.}\ }\textbf {\bibinfo {volume} {88}},\ \bibinfo {pages} {246102}
  (\bibinfo {year} {2002})}\BibitemShut {NoStop}%
\bibitem [{\citenamefont {{M. E. Leunissen {\sl et
  al.}}}(2007)}]{Leunissen2007}%
  \BibitemOpen
  \bibfield  {author} {\bibinfo {author} {\bibnamefont {{M. E. Leunissen {\sl
  et al.}}}},\ }\href@noop {} {\bibfield  {journal} {\bibinfo  {journal}
  {Proceedings of the National Academy of Sciences of the United States of
  America}\ }\textbf {\bibinfo {volume} {104}},\ \bibinfo {pages} {2585}
  (\bibinfo {year} {2007})}\BibitemShut {NoStop}%
\bibitem [{\citenamefont {{K. Masschaele, B. J. Park, E. M. Furst, J. Fransaer,
  and J. Vermant}}(2010)}]{Masschaele2010}%
  \BibitemOpen
  \bibfield  {author} {\bibinfo {author} {\bibnamefont {{K. Masschaele, B. J.
  Park, E. M. Furst, J. Fransaer, and J. Vermant}}},\ }\href@noop {} {\bibfield
   {journal} {\bibinfo  {journal} {Phys. Rev. Lett.}\ }\textbf {\bibinfo
  {volume} {105}},\ \bibinfo {pages} {048303} (\bibinfo {year}
  {2010})}\BibitemShut {NoStop}%
\bibitem [{\citenamefont {{J. Guzowski, M. Tasinkevych, and S.
  Dietrich}}(2011)}]{Guzowski2011}%
  \BibitemOpen
  \bibfield  {author} {\bibinfo {author} {\bibnamefont {{J. Guzowski, M.
  Tasinkevych, and S. Dietrich}}},\ }\href@noop {} {\bibfield  {journal}
  {\bibinfo  {journal} {Phys. Rev. E}\ }\textbf {\bibinfo {volume} {84}},\
  \bibinfo {pages} {031401} (\bibinfo {year} {2011})}\BibitemShut {NoStop}%
\bibitem [{\citenamefont {{M. Bowick, A. Cacciuto, D. R. Nelson, and A.
  Travesset}}(2002)}]{Bowick2002}%
  \BibitemOpen
  \bibfield  {author} {\bibinfo {author} {\bibnamefont {{M. Bowick, A.
  Cacciuto, D. R. Nelson, and A. Travesset}}},\ }\href@noop {} {\bibfield
  {journal} {\bibinfo  {journal} {Phys. Rev. Lett.}\ }\textbf {\bibinfo
  {volume} {89}},\ \bibinfo {pages} {185502} (\bibinfo {year}
  {2002})}\BibitemShut {NoStop}%
\bibitem [{\citenamefont {{M. K.-H. Kiessling}}(2009)}]{Kiessling2009}%
  \BibitemOpen
  \bibfield  {author} {\bibinfo {author} {\bibnamefont {{M. K.-H.
  Kiessling}}},\ }\href@noop {} {\bibfield  {journal} {\bibinfo  {journal} {J.
  Stat. Phys.}\ }\textbf {\bibinfo {volume} {136}},\ \bibinfo {pages} {275}
  (\bibinfo {year} {2009})}\BibitemShut {NoStop}%
\bibitem [{\citenamefont {{J. M. Caillol}}(1981)}]{Caillol1981}%
  \BibitemOpen
  \bibfield  {author} {\bibinfo {author} {\bibnamefont {{J. M. Caillol}}},\
  }\href@noop {} {\bibfield  {journal} {\bibinfo  {journal} {J. Physique
  Lettres}\ }\textbf {\bibinfo {volume} {42}},\ \bibinfo {pages} {L245}
  (\bibinfo {year} {1981})}\BibitemShut {NoStop}%
\bibitem [{\citenamefont {{A. R. Bausch {\sl et al.}}}(2003)}]{Bausch2003}%
  \BibitemOpen
  \bibfield  {author} {\bibinfo {author} {\bibnamefont {{A. R. Bausch {\sl et
  al.}}}},\ }\href@noop {} {\bibfield  {journal} {\bibinfo  {journal}
  {Science}\ }\textbf {\bibinfo {volume} {299}},\ \bibinfo {pages} {1716}
  (\bibinfo {year} {2003})}\BibitemShut {NoStop}%
\bibitem [{\citenamefont {{P. Lipowsky, M. J. Bowick, J. H. Meinke, D. R.
  Nelson, and A. R. Bausch}}(2005)}]{Lipowsky2005}%
  \BibitemOpen
  \bibfield  {author} {\bibinfo {author} {\bibnamefont {{P. Lipowsky, M. J.
  Bowick, J. H. Meinke, D. R. Nelson, and A. R. Bausch}}},\ }\href@noop {}
  {\bibfield  {journal} {\bibinfo  {journal} {Nature Mat.}\ }\textbf {\bibinfo
  {volume} {4}},\ \bibinfo {pages} {407} (\bibinfo {year} {2005})}\BibitemShut
  {NoStop}%
\bibitem [{\citenamefont {{M. J. Bowick, D. R. Nelson, and A.
  Travesset}}(2000)}]{Bowick2000}%
  \BibitemOpen
  \bibfield  {author} {\bibinfo {author} {\bibnamefont {{M. J. Bowick, D. R.
  Nelson, and A. Travesset}}},\ }\href@noop {} {\bibfield  {journal} {\bibinfo
  {journal} {Phys. Rev. B}\ }\textbf {\bibinfo {volume} {62}},\ \bibinfo
  {pages} {8738} (\bibinfo {year} {2000})}\BibitemShut {NoStop}%
\bibitem [{\citenamefont {{S. Sastry, D. S. Corti, P. G. Debenedetti, and F. H.
  Stillinger}}(1997)}]{Sastry1997}%
  \BibitemOpen
  \bibfield  {author} {\bibinfo {author} {\bibnamefont {{S. Sastry, D. S.
  Corti, P. G. Debenedetti, and F. H. Stillinger}}},\ }\href@noop {} {\bibfield
   {journal} {\bibinfo  {journal} {Phys. Rev. E}\ }\textbf {\bibinfo {volume}
  {56}},\ \bibinfo {pages} {5524} (\bibinfo {year} {1997})}\BibitemShut
  {NoStop}%
\bibitem [{\citenamefont {{M. J. Bowick and L. Giomi}}(2009)}]{Bowick2009}%
  \BibitemOpen
  \bibfield  {author} {\bibinfo {author} {\bibnamefont {{M. J. Bowick and L.
  Giomi}}},\ }\href@noop {} {\bibfield  {journal} {\bibinfo  {journal}
  {Advances in Physics}\ }\textbf {\bibinfo {volume} {58}},\ \bibinfo {pages}
  {449} (\bibinfo {year} {2009})}\BibitemShut {NoStop}%
\bibitem [{\citenamefont {{C. N. Likos}}(2001)}]{Likos2001}%
  \BibitemOpen
  \bibfield  {author} {\bibinfo {author} {\bibnamefont {{C. N. Likos}}},\
  }\href@noop {} {\bibfield  {journal} {\bibinfo  {journal} {Phys. Rep.}\
  }\textbf {\bibinfo {volume} {348}},\ \bibinfo {pages} {267} (\bibinfo {year}
  {2001})}\BibitemShut {NoStop}%
\bibitem [{\citenamefont {{E. Zaccarelli}}(2007)}]{Zaccarelli2007}%
  \BibitemOpen
  \bibfield  {author} {\bibinfo {author} {\bibnamefont {{E. Zaccarelli}}},\
  }\href@noop {} {\bibfield  {journal} {\bibinfo  {journal} {J. Phys.: Condens.
  Matter}\ }\textbf {\bibinfo {volume} {19}},\ \bibinfo {pages} {323101}
  (\bibinfo {year} {2007})}\BibitemShut {NoStop}%
\bibitem [{\citenamefont {{R. Fantoni, B. Jancovici, and G.
  T\'ellez}}(2003)}]{Fantoni2003}%
  \BibitemOpen
  \bibfield  {author} {\bibinfo {author} {\bibnamefont {{R. Fantoni, B.
  Jancovici, and G. T\'ellez}}},\ }\href@noop {} {\bibfield  {journal}
  {\bibinfo  {journal} {J. Stat. Phys.}\ }\textbf {\bibinfo {volume} {112}},\
  \bibinfo {pages} {27} (\bibinfo {year} {2003})}\BibitemShut {NoStop}%
\bibitem [{\citenamefont {{R. Fantoni and G. T\'ellez}}(2008)}]{Fantoni2008}%
  \BibitemOpen
  \bibfield  {author} {\bibinfo {author} {\bibnamefont {{R. Fantoni and G.
  T\'ellez}}},\ }\href@noop {} {\bibfield  {journal} {\bibinfo  {journal} {J.
  Stat. Phys.}\ }\textbf {\bibinfo {volume} {133}},\ \bibinfo {pages} {449}
  (\bibinfo {year} {2008})}\BibitemShut {NoStop}%
\bibitem [{\citenamefont {{F. Sausset, G. Tarjus, and D. R.
  Nelson}}(2010)}]{Sausset2010}%
  \BibitemOpen
  \bibfield  {author} {\bibinfo {author} {\bibnamefont {{F. Sausset, G. Tarjus,
  and D. R. Nelson}}},\ }\href@noop {} {\bibfield  {journal} {\bibinfo
  {journal} {Phys. Rev. E}\ }\textbf {\bibinfo {volume} {81}},\ \bibinfo
  {pages} {031504} (\bibinfo {year} {2010})}\BibitemShut {NoStop}%
\bibitem [{\citenamefont {{M. Ch\'avez-P\'aez {\sl et al.}}}(2003)}]{Paez2003}%
  \BibitemOpen
  \bibfield  {author} {\bibinfo {author} {\bibnamefont {{M. Ch\'avez-P\'aez
  {\sl et al.}}}},\ }\href@noop {} {\bibfield  {journal} {\bibinfo  {journal}
  {J. Chem. Phys.}\ }\textbf {\bibinfo {volume} {119}},\ \bibinfo {pages}
  {7461} (\bibinfo {year} {2003})}\BibitemShut {NoStop}%
\bibitem [{\citenamefont {{P. X. Viveros-M\'endez, J. M. M\'endez-Alcaraz, and
  P. Gonz\'alez-Mozuelos}}(2008)}]{Mendez2008}%
  \BibitemOpen
  \bibfield  {author} {\bibinfo {author} {\bibnamefont {{P. X.
  Viveros-M\'endez, J. M. M\'endez-Alcaraz, and P. Gonz\'alez-Mozuelos}}},\
  }\href@noop {} {\bibfield  {journal} {\bibinfo  {journal} {J. Chem. Phys.}\
  }\textbf {\bibinfo {volume} {128}},\ \bibinfo {pages} {014701} (\bibinfo
  {year} {2008})}\BibitemShut {NoStop}%
\bibitem [{\citenamefont {{S. Prestipino Giarritta, M. Ferrario, and P. V.
  Giaquinta}}(1992)}]{Prestipino1992}%
  \BibitemOpen
  \bibfield  {author} {\bibinfo {author} {\bibnamefont {{S. Prestipino
  Giarritta, M. Ferrario, and P. V. Giaquinta}}},\ }\href@noop {} {\bibfield
  {journal} {\bibinfo  {journal} {Physica A}\ }\textbf {\bibinfo {volume}
  {187}},\ \bibinfo {pages} {456} (\bibinfo {year} {1992})}\BibitemShut
  {NoStop}%
\bibitem [{\citenamefont {{S. Prestipino Giarritta, M. Ferrario, and P. V.
  Giaquinta}}(1993)}]{Prestipino1993}%
  \BibitemOpen
  \bibfield  {author} {\bibinfo {author} {\bibnamefont {{S. Prestipino
  Giarritta, M. Ferrario, and P. V. Giaquinta}}},\ }\href@noop {} {\bibfield
  {journal} {\bibinfo  {journal} {Physica A}\ }\textbf {\bibinfo {volume}
  {201}},\ \bibinfo {pages} {649} (\bibinfo {year} {1993})}\BibitemShut
  {NoStop}%
\bibitem [{\citenamefont {{J. W. O. Salari, G. T. Jemwa, H. M. Wyss, and B.
  Klumperman}}(2011)}]{Salari2010}%
  \BibitemOpen
  \bibfield  {author} {\bibinfo {author} {\bibnamefont {{J. W. O. Salari, G. T.
  Jemwa, H. M. Wyss, and B. Klumperman}}},\ }\href@noop {} {\bibfield
  {journal} {\bibinfo  {journal} {Soft Matter}\ }\textbf {\bibinfo {volume}
  {7}},\ \bibinfo {pages} {2033} (\bibinfo {year} {2011})}\BibitemShut
  {NoStop}%
\bibitem [{Sal()}]{Salari-Thesis}%
  \BibitemOpen
  \href@noop {} {}\bibinfo {note} {J. W. O. Salari, Ph.D. thesis, "Pickering
  emulsions, colloidosomes \& micro-encapsulation", Eindhoven University of
  Technology, 2011}\BibitemShut {NoStop}%
\bibitem [{\citenamefont {{C. E. Smith}}(2010)}]{Smith2010}%
  \BibitemOpen
  \bibfield  {author} {\bibinfo {author} {\bibnamefont {{C. E. Smith}}},\
  }\href@noop {} {\bibfield  {journal} {\bibinfo  {journal} {Journal of
  Physics: Conference Series}\ }\textbf {\bibinfo {volume} {237}},\ \bibinfo
  {pages} {012021} (\bibinfo {year} {2010})}\BibitemShut {NoStop}%
\bibitem [{\citenamefont {{L. Bruneau and S. De
  Bi\`evre}}(2002)}]{Bruneau2002}%
  \BibitemOpen
  \bibfield  {author} {\bibinfo {author} {\bibnamefont {{L. Bruneau and S. De
  Bi\`evre}}},\ }\href@noop {} {\bibfield  {journal} {\bibinfo  {journal}
  {Comm. Math. Phys.}\ }\textbf {\bibinfo {volume} {229}},\ \bibinfo {pages}
  {511} (\bibinfo {year} {2002})}\BibitemShut {NoStop}%
\bibitem [{\citenamefont {{S. U. Pickering}}(1907)}]{Pickering1907}%
  \BibitemOpen
  \bibfield  {author} {\bibinfo {author} {\bibnamefont {{S. U. Pickering}}},\
  }\href@noop {} {\bibfield  {journal} {\bibinfo  {journal} {J. Chem. Soc.}\
  }\textbf {\bibinfo {volume} {91}},\ \bibinfo {pages} {2001} (\bibinfo {year}
  {1907})}\BibitemShut {NoStop}%
\bibitem [{Note1()}]{Note1}%
  \BibitemOpen
  \bibinfo {note} {The right panel of the Fig. \ref {fig:2} shows clearly
  flattening of the particles on the inside of the capsules. We are convinced
  that this is an artifact of the sintering process. During this process
  flattening of the particles occurs, which we ascribe to particle deformation
  to reduce the contact area between particle and water.}\BibitemShut {Stop}%
\bibitem [{\citenamefont {{J. W. O. Salari, F. A. M. Leermakers, and B.
  Klumperman}}(2011)}]{Salari2011}%
  \BibitemOpen
  \bibfield  {author} {\bibinfo {author} {\bibnamefont {{J. W. O. Salari, F. A.
  M. Leermakers, and B. Klumperman}}},\ }\href@noop {} {\bibfield  {journal}
  {\bibinfo  {journal} {Langmuir}\ }\textbf {\bibinfo {volume} {27}},\ \bibinfo
  {pages} {6574} (\bibinfo {year} {2011})},\ \bibinfo {note} {and references
  therein}\BibitemShut {NoStop}%
\bibitem [{\citenamefont {{L. D. Landau and E. M. Lifshitz}}(1977)}]{Landau3}%
  \BibitemOpen
  \bibfield  {author} {\bibinfo {author} {\bibnamefont {{L. D. Landau and E. M.
  Lifshitz}}},\ }\href@noop {} {\emph {\bibinfo {title} {Quantum Mechanics.
  Non-relativistic Theory}}},\ \bibinfo {edition} {3rd}\ ed.,\ Vol.~\bibinfo
  {volume} {3}\ (\bibinfo  {publisher} {Pergamon Press},\ \bibinfo {year}
  {1977})\ \bibinfo {note} {course of Theoretical Physics. Section
  89.}\BibitemShut {Stop}%
\bibitem [{\citenamefont {{H. C. Hamaker}}(1937)}]{Hamaker1937}%
  \BibitemOpen
  \bibfield  {author} {\bibinfo {author} {\bibnamefont {{H. C. Hamaker}}},\
  }\href@noop {} {\bibfield  {journal} {\bibinfo  {journal} {Physica}\ }\textbf
  {\bibinfo {volume} {4}},\ \bibinfo {pages} {1058} (\bibinfo {year}
  {1937})}\BibitemShut {NoStop}%
\bibitem [{shs()}]{shs}%
  \BibitemOpen
  \href@noop {} {}\bibinfo {note} {Suppose that the particles had a
  sticky-hard-sphere interaction \cite{Fantoni2005a,*Fantoni2005b} the surface
  adhesion would still be canceled by the presence of the hairs.}\BibitemShut
  {Stop}%
\bibitem [{\citenamefont {{S. Jiang and S. Granick}}(2008)}]{Jiang2008}%
  \BibitemOpen
  \bibfield  {author} {\bibinfo {author} {\bibnamefont {{S. Jiang and S.
  Granick}}},\ }\href@noop {} {\bibfield  {journal} {\bibinfo  {journal}
  {Langmuir}\ }\textbf {\bibinfo {volume} {24}},\ \bibinfo {pages} {2438}
  (\bibinfo {year} {2008})}\BibitemShut {NoStop}%
\bibitem [{\citenamefont {{D. Gazzillo, A. Giacometti, R. Fantoni, and P.
  Sollich}}(2006)}]{Gazzillo2006}%
  \BibitemOpen
  \bibfield  {author} {\bibinfo {author} {\bibnamefont {{D. Gazzillo, A.
  Giacometti, R. Fantoni, and P. Sollich}}},\ }\href@noop {} {\bibfield
  {journal} {\bibinfo  {journal} {Phys. Rev. E}\ }\textbf {\bibinfo {volume}
  {74}},\ \bibinfo {pages} {051407} (\bibinfo {year} {2006})}\BibitemShut
  {NoStop}%
\bibitem [{\citenamefont {Allen}\ and\ \citenamefont
  {Tildesley}(1987)}]{Allen-Tildesley}%
  \BibitemOpen
  \bibfield  {author} {\bibinfo {author} {\bibfnamefont {M.~P.}\ \bibnamefont
  {Allen}}\ and\ \bibinfo {author} {\bibfnamefont {D.~J.}\ \bibnamefont
  {Tildesley}},\ }\href@noop {} {\emph {\bibinfo {title} {Computer Simulation
  of Liquids}}}\ (\bibinfo  {publisher} {Clarendon Press},\ \bibinfo {address}
  {Oxford},\ \bibinfo {year} {1987})\BibitemShut {NoStop}%
\bibitem [{\citenamefont {{D. Frenkel and B. Smit}}(1996)}]{Frenkel-Smit}%
  \BibitemOpen
  \bibfield  {author} {\bibinfo {author} {\bibnamefont {{D. Frenkel and B.
  Smit}}},\ }\href@noop {} {\emph {\bibinfo {title} {Understanding Molecular
  Simulation}}}\ (\bibinfo  {publisher} {Academic Press},\ \bibinfo {address}
  {San Diego},\ \bibinfo {year} {1996})\BibitemShut {NoStop}%
\bibitem [{\citenamefont {{E. R. Weeks and D. A. Weitz}}(2002)}]{Weeks2002}%
  \BibitemOpen
  \bibfield  {author} {\bibinfo {author} {\bibnamefont {{E. R. Weeks and D. A.
  Weitz}}},\ }\href@noop {} {\bibfield  {journal} {\bibinfo  {journal}
  {Chemical Physics}\ }\textbf {\bibinfo {volume} {284}},\ \bibinfo {pages}
  {361} (\bibinfo {year} {2002})}\BibitemShut {NoStop}%
\bibitem [{\citenamefont {{Grazyna Antczak and Gert Ehrlich}}(2010)}]{Ehrlich}%
  \BibitemOpen
  \bibfield  {author} {\bibinfo {author} {\bibnamefont {{Grazyna Antczak and
  Gert Ehrlich}}},\ }\href@noop {} {\emph {\bibinfo {title} {Surface Diffusion:
  Metals, Metal Atoms, and Clusters}}}\ (\bibinfo  {publisher} {Cambridge
  University Press},\ \bibinfo {address} {Cambridge},\ \bibinfo {year}
  {2010})\BibitemShut {NoStop}%
\bibitem [{\citenamefont {Fischer}(1958)}]{Fischer1958}%
  \BibitemOpen
  \bibfield  {author} {\bibinfo {author} {\bibfnamefont {E.~W.}\ \bibnamefont
  {Fischer}},\ }\href@noop {} {\bibfield  {journal} {\bibinfo  {journal}
  {Kolloid Z.}\ }\textbf {\bibinfo {volume} {160}},\ \bibinfo {pages} {120}
  (\bibinfo {year} {1958})}\BibitemShut {NoStop}%
\bibitem [{\citenamefont {{P. R. ten Wolde, M. J. Ruiz-Montero, and D.
  Frenkel}}(1996)}]{Wolde1996}%
  \BibitemOpen
  \bibfield  {author} {\bibinfo {author} {\bibnamefont {{P. R. ten Wolde, M. J.
  Ruiz-Montero, and D. Frenkel}}},\ }\href@noop {} {\bibfield  {journal}
  {\bibinfo  {journal} {J. Chem. Phys.}\ }\textbf {\bibinfo {volume} {104}},\
  \bibinfo {pages} {9932} (\bibinfo {year} {1996})}\BibitemShut {NoStop}%
\bibitem [{\citenamefont {{A. D. Law, D. M. A. Buzza, and T. S.
  Horozov}}(2011)}]{Law2011}%
  \BibitemOpen
  \bibfield  {author} {\bibinfo {author} {\bibnamefont {{A. D. Law, D. M. A.
  Buzza, and T. S. Horozov}}},\ }\href@noop {} {\bibfield  {journal} {\bibinfo
  {journal} {Phys. Rev. Lett.}\ }\textbf {\bibinfo {volume} {106}},\ \bibinfo
  {pages} {128302} (\bibinfo {year} {2011})}\BibitemShut {NoStop}%
\bibitem [{\citenamefont {{A. Santos, M. L. de Haro, and S. B.
  Yuste}}(2010)}]{Santos2010}%
  \BibitemOpen
  \bibfield  {author} {\bibinfo {author} {\bibnamefont {{A. Santos, M. L. de
  Haro, and S. B. Yuste}}},\ }\href@noop {} {\bibfield  {journal} {\bibinfo
  {journal} {J. Chem. Phys.}\ }\textbf {\bibinfo {volume} {132}},\ \bibinfo
  {pages} {204506} (\bibinfo {year} {2010})}\BibitemShut {NoStop}%
\bibitem [{\citenamefont {{R. Fantoni and A. Santos}}(2011)}]{Fantoni2011}%
  \BibitemOpen
  \bibfield  {author} {\bibinfo {author} {\bibnamefont {{R. Fantoni and A.
  Santos}}},\ }\href@noop {} {\bibfield  {journal} {\bibinfo  {journal} {Phys.
  Rev. E}\ }\textbf {\bibinfo {volume} {84}},\ \bibinfo {pages} {041201}
  (\bibinfo {year} {2011})}\BibitemShut {NoStop}%
\bibitem [{\citenamefont {{A. Santos, S. B. Yuste, M. L. de
  Haro}}(2011)}]{Santos2011}%
  \BibitemOpen
  \bibfield  {author} {\bibinfo {author} {\bibnamefont {{A. Santos, S. B.
  Yuste, M. L. de Haro}}},\ }\href@noop {} {\bibfield  {journal} {\bibinfo
  {journal} {J. Chem. Phys.}\ }\textbf {\bibinfo {volume} {135}},\ \bibinfo
  {pages} {181102} (\bibinfo {year} {2011})}\BibitemShut {NoStop}%
\bibitem [{\citenamefont {{R. Fantoni, A. Giacometti, A. Malijevsk\'y, and A.
  Santos}}(2009)}]{Fantoni09a}%
  \BibitemOpen
  \bibfield  {author} {\bibinfo {author} {\bibnamefont {{R. Fantoni, A.
  Giacometti, A. Malijevsk\'y, and A. Santos}}},\ }\href@noop {} {\bibfield
  {journal} {\bibinfo  {journal} {J. Chem. Phys.}\ }\textbf {\bibinfo {volume}
  {131}},\ \bibinfo {pages} {124106} (\bibinfo {year} {2009})}\BibitemShut
  {NoStop}%
\bibitem [{\citenamefont {{R. E. Goldstein, M. Polin, and I.
  Tuval}}(2011)}]{Goldstein2011}%
  \BibitemOpen
  \bibfield  {author} {\bibinfo {author} {\bibnamefont {{R. E. Goldstein, M.
  Polin, and I. Tuval}}},\ }\href@noop {} {\bibfield  {journal} {\bibinfo
  {journal} {Phys. Rev. Lett.}\ }\textbf {\bibinfo {volume} {107}},\ \bibinfo
  {pages} {148103} (\bibinfo {year} {2011})}\BibitemShut {NoStop}%
\bibitem [{\citenamefont {{T. L. Hill}}(1956)}]{Hill}%
  \BibitemOpen
  \bibfield  {author} {\bibinfo {author} {\bibnamefont {{T. L. Hill}}},\
  }\href@noop {} {\emph {\bibinfo {title} {Statistical Mechanics}}}\ (\bibinfo
  {publisher} {McGraw-Hill},\ \bibinfo {address} {New York},\ \bibinfo {year}
  {1956})\BibitemShut {NoStop}%
\bibitem [{\citenamefont {{M. H. Kalos and P. A.
  Whitlock}}(2008)}]{Kalos-Whitlock}%
  \BibitemOpen
  \bibfield  {author} {\bibinfo {author} {\bibnamefont {{M. H. Kalos and P. A.
  Whitlock}}},\ }\href@noop {} {\emph {\bibinfo {title} {Monte Carlo
  Methods}}}\ (\bibinfo  {publisher} {Wiley-Vch Verlag GmbH \& Co. KGaA},\
  \bibinfo {address} {Weinheim},\ \bibinfo {year} {2008})\BibitemShut {NoStop}%
\bibitem [{\citenamefont {{R. Fantoni, D. Gazzillo, and A.
  Giacometti}}(2005{\natexlab{a}})}]{Fantoni2005a}%
  \BibitemOpen
  \bibfield  {author} {\bibinfo {author} {\bibnamefont {{R. Fantoni, D.
  Gazzillo, and A. Giacometti}}},\ }\href@noop {} {\bibfield  {journal}
  {\bibinfo  {journal} {J. Chem. Phys.}\ }\textbf {\bibinfo {volume} {122}},\
  \bibinfo {pages} {034901} (\bibinfo {year} {2005}{\natexlab{a}})}\BibitemShut
  {NoStop}%
\bibitem [{\citenamefont {{R. Fantoni, D. Gazzillo, and A.
  Giacometti}}(2005{\natexlab{b}})}]{Fantoni2005b}%
  \BibitemOpen
  \bibfield  {author} {\bibinfo {author} {\bibnamefont {{R. Fantoni, D.
  Gazzillo, and A. Giacometti}}},\ }\href@noop {} {\bibfield  {journal}
  {\bibinfo  {journal} {Phys. Rev. E}\ }\textbf {\bibinfo {volume} {72}},\
  \bibinfo {pages} {011503} (\bibinfo {year} {2005}{\natexlab{b}})}\BibitemShut
  {NoStop}%
\end{thebibliography}%


%%%%%%%%%%%%%%%%%%%%%%%%%%%%%%%%%%%%%%%%%%%%%%%%%%%%%%%%%%%%%%%%%%%%%%%%%%%%%%%
%%%%%%%%%%%%%%%%%%%%%%%%%%%%%%%%%%%%%%%%%%%%%%%%%%%%%%%%%%%%%%%%%%%%%%%%%%%%%%%
%%%%%%%%%%%%%%%%%%%%%%%%%%%%%%%%%%%%%%%%%%%%%%%%%%%%%%%%%%%%%%%%%%%%%%%%%%%%%%%
\end{document}